\documentclass[twocolumn]{aastex631}

\AtBeginDocument{\renewenvironment{acknowledgments}{\section*{Acknowledgments}}{}}

\numlinesfalse
\AtBeginDocument{}
\usepackage{amsmath}
\usepackage{booktabs}

\renewcommand{\arraystretch}{1.12}

\newcommand{\JR}{J_R}
\newcommand{\Meff}{M_{\rm eff}}
\newcommand{\Msun}{\ensuremath{\mathrm{M_\odot}}}
\newcommand{\kms}{\mathrm{km\,s^{-1}}}
\newcommand{\kpckms}{\mathrm{kpc\,km\,s^{-1}}}
\newcommand{\dd}{\mathrm{d}}
\newcommand{\Ssc}{\Sigma_{\rm sc}}
\newcommand{\SGMC}{\Sigma_{\rm GMC}}
\newcommand{\NKS}{N_{\rm KS}}
\newcommand{\tcell}[2]{\parbox[t]{#1}{\raggedright #2}}

\shorttitle{Where the Disc is Heated}
\shortauthors{Ting \& Rix}

\begin{document}

\title{The Shape of the Vertical Action Distribution Locates the Scatterers that Heat the Galactic Disc}

\author[0000-0001-5082-9536]{Yuan-Sen Ting}
\affiliation{Department of Astronomy, The Ohio State University, 140 West 18th Avenue, Columbus, OH 43210, USA}
\affiliation{Center for Cosmology and AstroParticle Physics (CCAPP), The Ohio State University, Columbus, OH 43210, USA}
\affiliation{Max-Planck-Institut f\"ur Astronomie, K\"onigstuhl 17, D-69117 Heidelberg, Germany}

\author[0000-0003-4996-9069]{Hans-Walter Rix}
\affiliation{Max-Planck-Institut f\"ur Astronomie, K\"onigstuhl 17, D-69117 Heidelberg, Germany}

\begin{abstract}

The Milky Way's stellar disc is thicker than the cold gas layer from which its stars form. What scatters stars onto orbits with greater vertical motion remains unresolved. Scatterers could fill the disc volume, as bending waves or dark substructure would, or could be confined to the midplane, as giant molecular clouds are. Scattering from a midplane layer occurs only during the fast plane-crossing phase and becomes less effective as that speed increases, whereas a volume-filling perturbation remains effective near the slow turning points. We derive the distribution of vertical action this leaves behind, for any height distribution of scatterers. The geometry turns out to enter only through the logarithmic slope of the diffusivity in action, $D\propto J_z^{\,b}$, and solving the Fokker--Planck equation gives $p(J_z)\propto\exp[-(J_z/J_0)^{2-b}]$. Scatterers that fill the volume give $b=1$ and an exponential, a thin layer at the midplane gives $b=1/2$ and a sharper cutoff: the shape of $p(J_z)$ records where the scatterers sit, the growth of its scale how strongly they scatter. We fit this model to $7589$ low-$\alpha$ red clump stars of \citet{tingrix2019} between $5$ and $10$\,kpc and $2$ and $8$\,Gyr old, leaving the heating history free. This yields $b=0.51^{+0.06}_{-0.07}$, consistent with the thin-layer prediction but not the volume-filling one. Comparing the $2$--$4$\,Gyr heating amplitude with the present molecular surface density gives an effective scatterer mass of $2.7\times10^{6}\,\Msun$. Older stars have experienced more of the Galaxy's gas-richer past; correcting for that history brings all four age bins to $1.9$--$2.8\times10^{6}\,\Msun$, inside the range cloud catalogues and mass functions give. The Milky Way's disc is heated near the plane, by an evolving population of objects of giant-molecular-cloud mass.

\end{abstract}


\section{Introduction}
\label{sec:introduction}

The random motions of disc stars, in particular their vertical motions, grow with age \citep[e.g.][]{wielen1977,holmberg2009,aumerbinney2009}. The scattering mechanism has been debated since \citet{spitzer1951,spitzer1953} proposed massive interstellar clouds as scatterers, before such clouds had even been observed \citep{jenkinsbinney1990}. The candidates put forward since divide into two classes, according to where along a star's orbit they act. Molecular clouds sit in a thin layer about the Galactic midplane \citep{nakanishisofue2006,miville2017}, so a star meets them only while crossing it.

The other class of models has the scatterers spread through the disc volume or beyond it, where they act on a disc star wherever it happens to be. Large-scale spiral structure is the oldest of them \citep{barbaniswoltjer1967,carlberg1985,desimone2004}. The Galactic bar, infalling satellites, and self-gravitating bending and breathing waves have been added since \citep[e.g.][]{jenkins1992,weinberg2001,saha2010,widrow2012,fouvry2017,tremaine2023}. Dark scatterers have also been proposed. \citet{laceyostriker1985} argued that haloes made of black holes could supply the observed heating if the holes weighed of order $10^{6}\,\Msun$.

Both classes are viable, and the measurements brought to bear so far have not separated them. Molecular clouds deliver about the observed amount of heating, provided their masses lie near the top of the range the observations then allowed \citep{lacey1984,villumsen1985}. Transient spirals reproduce the age--velocity dispersion relation together with the moving groups seen in local velocity space \citep{carlberg1985,desimone2004}. Simulations recover the observed heating with substantially different scatterer populations \citep{hanninenflynn2002,aumer2016,grand2016}. The shape of the velocity ellipsoid was proposed as a discriminant by \citet{binneylacey1988} and \citet{jenkinsbinney1990}: clouds turn in-plane motion into vertical motion efficiently and large-scale structure does not, so the ratio $\sigma_z/\sigma_R$ ought to record which class is at work. Neither class reproduced the observed ratio on its own, and it was matched instead by letting both act, with spirals supplying most of the in-plane energy and clouds redirecting part of it upward.

The main obstacle to discriminating between these model classes has been the ``observable'' itself: the velocity \emph{dispersion} is only a one-number summary at each age of the kinematics stars are born with, the abundance and history of the perturbers, the restoring potential, and the way the heating depends on the orbit. This leaves serious degeneracies about the scatterers, as the heating rate constrains essentially the product of their surface density and their effective mass, so many light perturbers and few heavy ones are interchangeable \citep{lacey1984}. Better data \citep[e.g.][]{nordstrom2004,mackereth2019,sharma2021} have improved heating rate estimates without settling the mechanism. Demonstrating that a given perturber can heat the disc is therefore no longer decisive, since its abundance can be chosen to match the observed amount whatever the perturber is and wherever along the orbit it acts.

The vertical \emph{action} may be the best diagnostic (among measures of vertical motion) to differentiate the two model classes. It is the phase-space area enclosed by one vertical oscillation, so it labels the orbit rather than the instantaneous velocity. It is also an adiabatic invariant, conserved when the background potential changes slowly compared with a vertical period \citep{binneytremaine2008}. The mass of the disc grew over the lifetimes of these stars, deepening the vertical potential and raising the vertical energy of every star, including stars that were never scattered at all. A distribution of vertical energies therefore mixes heating with the growth of the disc, whereas the action is insensitive to the second and records the first.

The distribution has more to say than its mean. Heating is a diffusion in the space of orbits. \citet{binneylacey1988} showed that the Fokker--Planck equation describing it takes a particularly simple form in action variables, with the drift fixed by the divergence of the diffusion tensor. That structure constrains the distribution of vertical action within a coeval population. Its shape is set by how steeply the diffusivity grows with the action rather than free to take any form, and one exponent controls both that shape and the rate at which the population spreads.

\citet{binneylacey1988} posed this question once before and took it as far as the data then allowed. Requiring that the observed vertical velocity distribution be Gaussian and stay Gaussian, they found that the diffusivity must grow in proportion to the vertical energy, and that scattering off clouds does not supply it.

Vertical actions for thousands of stars spanning several kiloparsecs, each with an age, change what can be asked. The calculations of \citet{binneylacey1988} fixed the geometry at the outset, with the vertical case assuming that every encounter happens as the star crosses the plane, and each returned a diffusion coefficient for its own case alone. The exponent can instead be treated as a free parameter and fitted, but only with a theory in which it is free.

Here we derive instead one expression that holds for any height distribution of scatterers, so the two ideal cases and a layer of real thickness all follow from it rather than from separate calculations. It separates what the shape of the distribution says from what its growth says. The shape of a single-age population is set by the geometry alone, so where the scatterers sit can be measured without assuming anything about how the heating rate evolved. The growth carries their strength, which for a population of clouds measures an effective mass.

The sample needed for both measurements already exists. In \citet{tingrix2019} we measured how the mean vertical action of the low-$\alpha$ disc grows with age and radius, from APOGEE red clump stars \citep{majewski2017} with \textit{Gaia} astrometry \citep{gaia2018}, forward-modelling the strong selection in vertical action. What that work fitted was the amount of heating, not the shape of the distribution it leaves behind, so the mechanism behind it could only be argued qualitatively. Here we return to the same stars, the same actions and the same selection model, cut to a range in radius and age where the actions are reliable and heating dominates, and fit the shape at each age rather than the mean.

\begin{figure*}[t]
\centering
\includegraphics[width=0.86\textwidth]{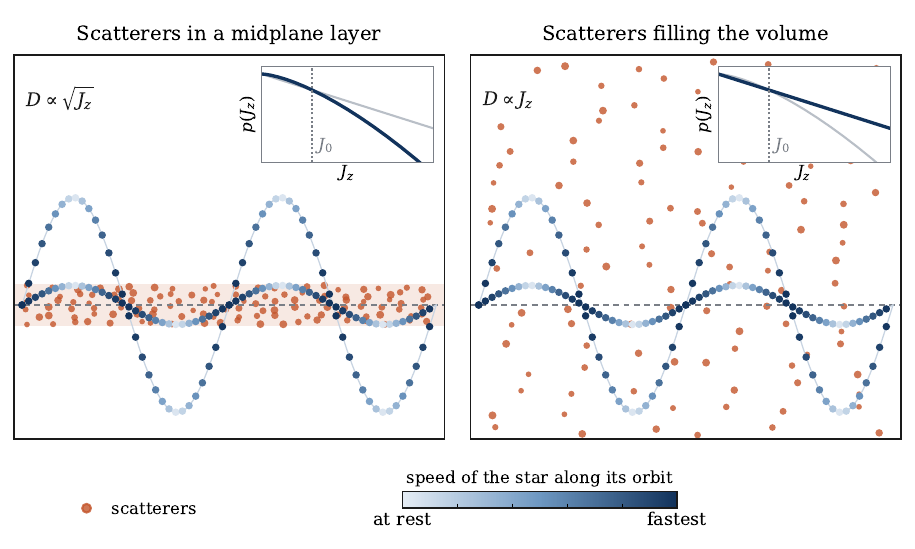}
\caption{Schematic of the two limiting scattering geometries and the distribution each produces. Each panel shows the same star on a low orbit and on one reaching much higher, among the same number of scatterers arranged two ways. The dots along each orbit are spaced at equal intervals of time, so they crowd where the star lingers and spread apart where it hurries. Each inset gives the distribution of vertical action within a coeval population, on a logarithmic axis, in colour for that panel's geometry and in grey for the other. Both are drawn with the same scale, set where the density has fallen by one $e$-folding, so only the tail differs. In the left panel, confinement means the scatterers are met only during the fast crossings, so the higher a star climbs the less of its time it spends among them, the heating falls behind the orbit, and the distribution is cut off more sharply than an exponential. In the right panel, scatterers filling the volume are met everywhere, so the heating keeps pace, the high orbit is the low one enlarged, and the distribution stays exponential.}
\label{fig:mechanism}
\end{figure*}

Section~\ref{sec:physical} builds the physical picture, and Section~\ref{sec:theory} turns it into a calculation, from a single encounter through to the distribution of a coeval population and the exponent each geometry implies. Section~\ref{sec:measurement} describes the sample, carries the strong selection in vertical action through the fit, and reports the exponent against what a real molecular layer predicts. Section~\ref{sec:amplitude} turns the amplitude of the same heating into an effective scatterer mass. Section~\ref{sec:discussion} sets the two results against the cloud catalogues and against the earlier determinations, and sets out the approximations they rest on. The appendices carry the derivations, test the estimator against known geometries, and relax the idealisations the main text works under.

\begin{table*}[t]
\centering
\caption{Symbols used throughout. Actions are in $\kpckms$, masses in \Msun, ages and times in Gyr, and frequencies in $\kms\,{\rm kpc}^{-1}$.}
\label{tab:symbols}
\small
\hspace*{-27pt}\begin{tabular}{lll}
\toprule
Role & Symbol & Definition\\
\midrule
Orbit & $J_z$, $Z$, $\nu$, $\Phi$ & \tcell{0.67\textwidth}{Vertical action $J_z=(2\pi)^{-1}\oint v_z\,\dd z$, the phase-space area enclosed by one oscillation of a star with vertical velocity $v_z$. The orbit reaches a height $Z=\sqrt{2J_z/\nu}$, and the vertical potential $\Phi$ fixes the frequency through $\Phi(z)-\Phi(0)\simeq\tfrac12\nu^2z^2$}\\[2pt]
& $\JR$, $\kappa$, $E_z$, $E_R$ & \tcell{0.67\textwidth}{Radial action and epicyclic frequency, the in-plane counterparts, and the two energies $E_z=\nu J_z$ and $E_R=\kappa\JR$}\\
\midrule
Scattering & $\Delta v_z$ & \tcell{0.67\textwidth}{Vertical velocity change delivered by one encounter, unbiased, so that $\langle\Delta v_z\rangle=0$}\\[2pt]
& $q(z)$, $q_0$, $Q$, $\hat g$, $h$ & \tcell{0.67\textwidth}{Mean square velocity kick per unit time at height $z$, $q=\langle(\Delta v_z)^2\rangle/\Delta t$. Scatterers filling the volume make it the constant $q_0$, a thin layer makes it a column $Q=\int q\,\dd z$. The unit-normalised shape in $q(z)=Q\,\hat g(z)$ has for its width the scattering layer's scale height $h$}\\[2pt]
& $D(J_z)$, $b$, $D_0$ & \tcell{0.67\textwidth}{Diffusivity in action, its logarithmic slope $b=\dd\ln D/\dd\ln J_z$, and its amplitude in $D=D_0J_z^{\,b}$}\\[2pt]
& $\mathcal{A}$ & \tcell{0.67\textwidth}{Drift coefficient in action, not independent but fixed by the diffusivity through $\mathcal{A}=\tfrac12\,\dd D/\dd J_z$}\\
\midrule
Population & $p(J_z)$, $J_0$, $s$, $\tau$, $t$ & \tcell{0.67\textwidth}{Distribution of vertical action in a coeval population, $p(J_z)\propto\exp[-(J_z/J_0)^{2-b}]$, with scale $J_0$, tail exponent $s=2-b$, and stellar age $\tau$. The diffusion clock has run for $t=\tau$ at a constant heating rate}\\[2pt]
& $J_{\rm b}$, $\sigma_z$, $\sigma_{\rm b}$ & \tcell{0.67\textwidth}{Scale of the distribution a population is born with, and the vertical velocity dispersion. At $b=1$ the solution is isothermal, with $J_0=\sigma_z^{2}/\nu$. The birth dispersion $\sigma_{\rm b}$ is its counterpart at formation, $\sigma_{\rm b}^{2}=\nu J_{\rm b}$}\\
\midrule
Measurement & $S_i$, $J_{{\rm t},i}$, $\xi_i$, $c_i$ & \tcell{0.67\textwidth}{Selection function of star $i$, unity below the transition action $J_{{\rm t},i}$ and $c_iJ_z^{-\xi_i}$ above, with all three polynomials in the star's radius}\\[2pt]
& $R_{\rm eff}$, $L$, $\alpha$ & \tcell{0.67\textwidth}{Radius at which a star was heated, the average of its current and birth radii, and the free scale history $\log J_0(\tau,R)=L(\log\tau)+\alpha\,(R-8\,{\rm kpc})$, with $L$ piecewise linear through four knots and $\alpha$ the log-slope in radius}\\
\midrule
Scatterers & $\Ssc$, $\Meff$ & \tcell{0.67\textwidth}{Surface density of the scatterers, $\Ssc=\int MN(M)\,\dd M$, and their mass-weighted mean mass, $\Meff=\int M^2N\,\dd M/\int MN\,\dd M$, for a mass function $N(M)$ per unit area}\\[2pt]
& $\SGMC$, $X_{\rm CO}$, $\ln\Lambda$, $V$ & \tcell{0.67\textwidth}{Molecular surface density from the cloud catalogue, which rests on the CO-to-$\rm H_2$ conversion factor $X_{\rm CO}$, the Coulomb logarithm $\ln(\varpi_{\max}/\varpi_{\min})$ over impact parameter $\varpi$, and the relative speed of star and cloud}\\
\midrule
History & $\Sigma_{\rm SFR}$, $\NKS$ & \tcell{0.67\textwidth}{Star-formation-rate surface density and the slope of the molecular Kennicutt--Schmidt relation $\Sigma_{\rm SFR}\propto\Ssc^{\NKS}$, which converts a star-formation history into a history of $\Ssc$}\\[2pt]
& $t_{\rm lb}$, $t_{\rm eff}$, $T$ & \tcell{0.67\textwidth}{Lookback time measured from the present along a cohort, the duration of heating at the present-day $\Ssc$ that would accumulate the same amount, and the e-folding time of the exponentially declining star-formation history}\\
\bottomrule
\end{tabular}
\end{table*}

\section{Why the Location of the Scatterers Sets the Heating}
\label{sec:physical}

Consider a star oscillating vertically through the disc, rising, slowing, turning and falling back, and perturbed at random along the way by the gravity of the masses it passes. Its orbit grows as a result. How fast it grows depends on how strong the perturbations are. How that growth changes as the orbit gets larger depends instead on where along the orbit they act. Two properties of the orbit govern it, and they act in opposite directions.

The first is that an impulse changes the energy in proportion to the speed at which it arrives. The work done is the product of the impulse and the velocity, so a given impulse is far more effective on a star moving quickly than on one nearly at rest. A star crosses the midplane at its greatest speed and reaches its turning points at rest. An encounter near the plane therefore changes the orbit substantially, and one near the turning point changes it hardly at all.

The second is that the star does not spend equal time at all heights. It lingers where it moves slowly, near the turning points, and passes quickly through the plane. The heights at which an impulse is most effective are precisely those the star occupies most briefly.

Which of the two dominates depends on where the scatterers lie. Figure~\ref{fig:mechanism} shows the two limiting cases side by side.

If the scatterers fill the volume the orbit sweeps, the star is among them at every point of it, and climbing higher does not carry it away from them. Only the strength of the kicks matters then, and a kick is the more effective the faster the star is moving, so an orbit that is already wide is heated faster than a narrow one, and faster in exact proportion. Growth in proportion to what is already there carries no preferred size, and the vertical actions of a coeval population spread into an exponential.

If instead the scatterers are confined to a thin layer about the plane, the star meets them only during its crossings. The impulses are large, because the crossing is where the star moves fastest. The crossing is brief for the same reason, since the time spent within a layer of fixed thickness falls as the crossing speed rises. The extra speed and the shorter stay cancel one against the other, and the heating grows only as the square root of the action rather than in proportion to it.

A star heated early climbs to an orbit that spends proportionally less of its time within the layer, so its subsequent heating slows. The layer thickness is a fixed length the orbit is measured against, and once the star has climbed past it the heating no longer keeps pace with the orbit, so the distribution is cut off more sharply than an exponential. Scatterers filling the volume offer no such length, the heating keeps pace at every height, and the distribution stays exponential. The difference is most pronounced among the stars that have been heated the most, which is why the distinction is carried by the tail of the distribution and not by its mean.

\section{Diffusion in Vertical Action}
\label{sec:theory}

The two competing effects of Section~\ref{sec:physical} become a calculation once they are written in the vertical action. Encounters are many and each is small, so the action performs a random walk, and what is wanted is not the fate of one star but the distribution of $J_z$ across a population of a single age. Two things are assumed. An encounter is brief compared with a vertical period, and the vertical action diffuses independently of the radial action. The perturbers then act as an unbiased random field, and nothing about them enters but their distribution in height. Table~\ref{tab:symbols} defines the symbols used throughout.

\subsection{The vertical action and a single encounter}

A diffusion equation needs a coordinate to diffuse in, and the choice decides how simple everything after it will be. The natural variable is not the height, nor the vertical velocity, nor the vertical energy, but the action
\begin{equation}
J_z=\frac{1}{2\pi}\oint v_z\,\dd z ,
\end{equation}
the phase-space area enclosed by one vertical oscillation. It does not depend on orbital phase, so it labels an orbit rather than an instant within one, which matters because each star is observed at whatever phase it happens to be in. It is also an adiabatic invariant, so it separates heating from the growth of the disc. Everything about the perturbers enters through one profile, the local rate at which they randomise the vertical velocity,
\begin{equation}
q(z)\equiv\lim_{\Delta t\to0}
\frac{\big\langle(\Delta v_z)^{2}\big\rangle}{\Delta t}\bigg|_{z} ,
\qquad \langle\Delta v_z\rangle=0 .
\end{equation}
The kicks are unbiased and symmetric, which is what it means for the perturbers to be an external random field rather than a systematic force.

In the impulse limit an encounter changes $v_z$ while leaving $z$ fixed. The vertical energy is $E_z=\tfrac12v_z^{2}+\Phi(z)$, and at fixed $z$ the potential term is unchanged, so
\begin{equation}
\Delta E_z=\tfrac12\big(v_z+\Delta v_z\big)^{2}-\tfrac12v_z^{2}
=v_z\,\Delta v_z+\tfrac12(\Delta v_z)^{2} ,
\label{eq:dE}
\end{equation}
exact for a kick of any size.

Near the plane the restoring force is close to linear, and writing $\Phi(z)-\Phi(0)\simeq\tfrac12\nu^{2}z^{2}$ defines the vertical frequency $\nu$. The total midplane density fixes it through $\nu^{2}\simeq4\pi G\rho_{\rm tot}(0)$, so stars, gas and dark matter contribute together. The motion is then a harmonic oscillation of amplitude $Z$,
\begin{equation}
z=Z\sin\nu t ,\qquad v_z=\nu Z\cos\nu t ,\qquad E_z=\tfrac12\nu^{2}Z^{2} .
\label{eq:harmonic}
\end{equation}
The ellipse this traces in the $(z,v_z)$ plane has semi-axes $Z$ and $\nu Z$, so its area is $\pi\nu Z^{2}$ and the action, that area divided by $2\pi$, is
\begin{equation}
\begin{aligned}
J_z&=\tfrac12\nu Z^{2}=\frac{E_z}{\nu} ,\\[2pt]
Z&=\sqrt{\frac{2J_z}{\nu}} ,\qquad
|v_z|=\nu\sqrt{Z^{2}-z^{2}} ,
\end{aligned}
\label{eq:orbit}
\end{equation}
the second by inverting the first and the third by eliminating $t$. The useful property of the harmonic limit is that $\nu$ does not depend on the amplitude, so large orbits and small ones take the same time and a star's vertical period does not change as it is heated. An impulse $\Delta v_z$ delivered while the star is moving at $v_z$ then changes the action by
\begin{equation}
\Delta J_z=\frac{v_z\,\Delta v_z}{\nu}+\frac{(\Delta v_z)^{2}}{2\nu} .
\label{eq:kick}
\end{equation}

The first term carries the sign of the kick, so it averages to zero while its square does not, and it is the source of the diffusion. The proportionality to $v_z$ matters. A kick of given size changes the action in proportion to how fast the star was moving when it arrived. A kick delivered at a turning point does nothing at first order, while one delivered at the midplane is maximally effective. The second term is positive whatever the sign of the kick, and is the systematic heating. It is not independent physics but the second-order part of the same kick, which is why no separate heating term has to be added by hand.

\subsection{The jump moments and the diffusivity}

One encounter is not the object of interest. The object is how the distribution moves under many of them, and a Fokker--Planck equation supplies it,
\begin{equation}
\frac{\partial p}{\partial t}
=-\frac{\partial}{\partial J_z}\big[\mathcal{A}\,p\big]
+\frac12\frac{\partial^{2}}{\partial J_z^{2}}\big[D\,p\big] ,
\label{eq:fpgeneral}
\end{equation}
fixed by two coefficients, the mean change per unit time and the variance per unit time. Appendix~\ref{app:fp} obtains it from the master equation for the encounter kicks, by truncating the Kramers--Moyal expansion at second order. Both coefficients follow from equation~\eqref{eq:kick}. Averaging it at fixed height, the first term drops because $\langle\Delta v_z\rangle=0$ and the second gives $q(z)$ per unit time, so
\begin{equation}
\mathcal{A}\big|_z=\frac{q(z)}{2\nu} ,
\qquad
D\big|_z=\frac{v_z^{2}\,q(z)}{\nu^{2}} .
\end{equation}
The variance follows from squaring equation~\eqref{eq:kick} before averaging. That leaves $v_z^{2}(\Delta v_z)^{2}/\nu^{2}$, a cross term odd in $\Delta v_z$ which vanishes on averaging, and a term in $(\Delta v_z)^{4}$ which falls faster than $\Delta t$ and drops in the limit. The drift comes entirely from the term a single kick contributes regardless of its sign.

Both coefficients hold at a fixed height, and what the star feels is their average around its orbit. Since $\dd t=\dd z/|v_z|$, a time average over one period becomes an integral over height,
\begin{equation}
\langle f\rangle_t=\frac{1}{T}\oint f\,\dd t
=\frac{2}{T}\int_{-Z}^{Z}f(z)\,\frac{\dd z}{|v_z|} ,
\qquad T=\frac{2\pi}{\nu} ,
\label{eq:dwell}
\end{equation}
the factor of two because every height is visited twice per period. The weight $\dd z/|v_z|$ is the dwell time, the time spent per unit height, and it makes the star linger at its turning points and pass quickly through the plane. Applying it to the variance,
\begin{equation}
\big\langle v_z^{2}q\big\rangle_t
=\frac{2}{T}\int_{-Z}^{Z}\nu^{2}\big(Z^{2}-z^{2}\big)\,q(z)\,
\frac{\dd z}{\nu\sqrt{Z^{2}-z^{2}}} .
\end{equation}
Since $D=\langle v_z^{2}q\rangle_t/\nu^{2}$, the factors of $\nu$ cancel between the kick and the dwell time, leaving
\begin{equation}
D(J_z)=\frac{1}{\pi}\int_{-Z}^{Z}q(z)\,\sqrt{Z^{2}-z^{2}}\;\dd z ,
\qquad Z=\sqrt{\frac{2J_z}{\nu}} .
\label{eq:master}
\end{equation}

The diffusivity is the scattering profile smeared with a semicircular window whose radius is the height the orbit reaches. One power of $|v_z|$ came from the kick and one inverse power from the dwell time, and what survives is their product. That product is the window itself, because $\nu\sqrt{Z^{2}-z^{2}}$ is the speed at height $z$. The window's peak and its width both grow as $\sqrt{J_z}$, so what is left to settle is how many of those powers the diffusivity takes up, and that is fixed by how much of the window the scatterers cover. Equation~\eqref{eq:master} converts a statement about where the scatterers are into a statement about how the heating grows with orbit size.

\subsection{The drift fixed by the diffusivity}

The drift has been carried along but not yet used, and it is not independent of the diffusivity. Differentiating equation~\eqref{eq:master} under the integral, with $\dd(Z^{2})/\dd J_z=2/\nu$,
\begin{equation}
\frac{\dd D}{\dd J_z}=\frac{1}{\pi\nu}\int_{-Z}^{Z}
\frac{q(z)}{\sqrt{Z^{2}-z^{2}}}\;\dd z ,
\end{equation}
the boundary term vanishing because the integrand does. Differentiating the window has turned it into the dwell-time weight, since $|v_z|=\nu\sqrt{Z^{2}-z^{2}}$ makes the right-hand side $\langle q\rangle_t/\nu$ in the average of equation~\eqref{eq:dwell}. Orbit-averaging $\mathcal{A}\big|_z=q/2\nu$ gives $\mathcal{A}=\langle q\rangle_t/2\nu$, so the two are the same quantity up to a factor,
\begin{equation}
\mathcal{A}=\frac12\,\frac{\dd D}{\dd J_z} ,
\label{eq:drift}
\end{equation}
the one-dimensional case of the relation \citet{binneylacey1988} established between the drift vector and the divergence of the diffusion tensor. It is not an assumption but a consequence of the kick being unbiased and depending on position alone. It holds for any $q(z)$, and it is a fluctuation--dissipation relation, so the problem carries one unknown function rather than two.

The reason is clearest in the velocity variable. Since $q$ depends on height alone, a drift could only arise from the kick strength changing with the star's own velocity, and a passing cloud does not know how fast the star is moving. The velocity-space flux is therefore purely down the gradient, and two neighbouring populations of equal density exchange stars and drive no net current.

Substituted into equation~\eqref{eq:fpgeneral}, this makes the two terms collapse into one. Expanding the second derivative gives $\tfrac12\partial_{J_z}[D'p+Dp']$, and the drift term $-\partial_{J_z}(\mathcal{A}p)$ removes the first of these exactly, so what remains is purely conservative,
\begin{equation}
\frac{\partial p}{\partial t}
=\frac12\frac{\partial}{\partial J_z}\left[D\frac{\partial p}{\partial J_z}\right] ,
\label{eq:fp}
\end{equation}
with no heating term at all. The population heats all the same. An action cannot be negative, so nothing crosses the origin. The flux, meaning the rate at which probability passes a given action, therefore vanishes there. The growth of $\langle J_z\rangle$ is then diffusion against a reflecting wall, which pushes the distribution outwards for the same reason that a random walk started at a wall drifts away from it. Heating requires no drift because the boundary supplies the asymmetry.

\subsection{The distribution of a coeval population}

Equation~\eqref{eq:fp} is controlled by one function. Take it to be a power law,
\begin{equation}
D=D_0J_z^{\,b} ,\qquad b=\frac{\dd\ln D}{\dd\ln J_z} .
\label{eq:powerlaw}
\end{equation}
The form is not a claim that the heating is independent of orbit size, since that dependence is what $b$ measures. It is the weaker claim that the dependence carries no scale of its own.

What makes that hold is that the scattering profile carries no length. A uniform $q$ has none, and an infinitely thin layer has none either, since a delta function has no width. Give the layer a finite thickness and a length enters, the power law survives only locally, and the exponent drifts with action, as Appendix~\ref{app:thickness} sets out. The exponent itself is not assumed but follows from the geometry.

The diffusivity $D_0$ is taken to be constant in time and the population to be born cold. Both simplifications move the scale of the answer rather than its shape, which is what the measurement later exploits by leaving the scale free.

Equation~\eqref{eq:fp} is solved through its moments, by the technique that solves the heat equation. Multiplying by $J_z^{\,n}$ and integrating by parts twice moves both derivatives off $p$ and onto the weight, and with the diffusivity no longer constant,
\begin{equation}
\frac{\dd}{\dd t}\big\langle J_z^{\,n}\big\rangle
=\frac{D_0}{2}\,n\,(n+b-1)\,\big\langle J_z^{\,n-(2-b)}\big\rangle .
\label{eq:ladder}
\end{equation}
The step is no longer two but $2-b$, and this is where the tail exponent comes from.

The ladder can be checked at the one value of $b$ where the answer is already known. At $b=0$ the diffusivity is constant, equation~\eqref{eq:fp} is the heat equation, and the ladder reduces to $\dd\langle J_z^{\,n}\rangle/\dd t=\tfrac12D_0\,n(n-1)\langle J_z^{\,n-2}\rangle$, each moment driven by the one two steps below it. Its $n=2$ member,
\begin{equation}
\frac{\dd}{\dd t}\big\langle J_z^{2}\big\rangle
=(1+b)\,D_0\big\langle J_z^{\,b}\big\rangle ,
\label{eq:second}
\end{equation}
returns the constant $D_0$ of the heat equation at $b=0$, so the only new ingredient is a diffusivity that couples a moment to a lower one instead of to a constant. That the hierarchy closes at all is not generic. For an arbitrary $D(J_z)$ the derivative of one moment involves $\langle J_z^{\,n-2}D\rangle$, which is not a moment of the family, and the method fails. It is the power law that makes each rung depend on exactly one rung below. The rungs $n=0,\,(2-b),\,2(2-b),\dots$ are spaced by $2-b$, so the ladder can be climbed from the bottom.

Relabelling the rungs by an integer index makes the recursion a single step. Writing $M_m\equiv\langle J_z^{\,m(2-b)}\rangle$, and taking $b<2$ so that the steps lead downwards rather than upwards,
\begin{equation}
\frac{\dd M_m}{\dd t}
=\frac{D_0}{2}\,m(2-b)\big[1+(m-1)(2-b)\big]\,M_{m-1} ,
\end{equation}
with $M_0=1$ by normalisation. A population born cold has $M_m(0)=0$ for $m\ge1$, so each step raises the power of $t$ by one and $M_m=t^{m}\prod_{k=1}^{m}c_k/m!$, with $c_k$ the coefficient above. The brackets telescope into a ratio of gamma functions,
\begin{equation}
M_m=\left[\frac{(2-b)^{2}D_0t}{2}\right]^{m}
\frac{\Gamma\!\left(m+\frac{1}{2-b}\right)}{\Gamma\!\left(\frac{1}{2-b}\right)} .
\label{eq:moments}
\end{equation}
Every moment is now known, and the bracket is the same at every rung. It defines the scale
\begin{equation}
J_0(t)=\left[\frac{(2-b)^{2}D_0t}{2}\right]^{1/(2-b)} ,
\label{eq:clock}
\end{equation}
the clock through which the whole heating history enters the problem and through which nothing else does.

The equation has handed over every moment, so the argument can now run in reverse, with the distribution that carries them as the answer. Setting $u=(J_z/J_0)^{2-b}$ turns equation~\eqref{eq:moments} into
\begin{equation}
\big\langle u^{m}\big\rangle
=\frac{\Gamma\!\left(m+\frac{1}{2-b}\right)}{\Gamma\!\left(\frac{1}{2-b}\right)} ,
\end{equation}
which are the moments of a gamma variable of shape $1/(2-b)$, so $u$ is that variable. Changing back with $\dd u/\dd J_z=(2-b)J_z^{1-b}/J_0^{2-b}$ and normalising,
\begin{equation}
p(J_z)=\frac{2-b}{\Gamma\!\left(\frac{1}{2-b}\right)J_0}
\exp\!\left[-\left(\frac{J_z}{J_0}\right)^{2-b}\right] .
\label{eq:solution}
\end{equation}
At $b=0$, where the ladder was checked against the heat equation, this is a Gaussian in the action. A diffusivity that grows with the action returns a stretched exponential instead, and the stretching index is the rung spacing. Recovering a distribution from its moments needs the moment problem to be determinate. Appendix~\ref{app:similarity} reaches equation~\eqref{eq:solution} a second way, by reducing the partial differential equation directly, and it sets out what \citet{binneylacey1988} obtained by the same route. The mathematics is not new. \citet{malacarne2001} give the same Green function and the same scale for a power-law diffusivity, as the linear limit of the nonlinear diffusion equation they treat, their $\theta$ standing for our $-b$. What the derivation above adds is the meaning of $b$, which equation~\eqref{eq:master} fixes from the height distribution of the scatterers.

Setting $J_z=J_0$ makes the exponential factor $e^{-1}$, so it is the action at which the distribution has fallen by one $e$-folding, the analogue of $\sigma$ for a Gaussian. It is not the mean, since the two differ by a ratio of gamma functions, which equals unity for an exponential and is smaller for steeper tails. There is a second reading at $b=1$, where the solution is the classical isothermal one. A Boltzmann distribution in vertical energy is $p\propto\exp(-E_z/\sigma_z^{2})$, and $E_z=\nu J_z$, so $J_0=\sigma_z^{2}/\nu$, the vertical temperature of the population divided by the restoring frequency.

Two features of equation~\eqref{eq:solution} matter for the measurement. The shape depends only on $b$, and the entire heating history enters only through $J_0$. So the exponent can be measured without assuming any form for that history, provided $J_0$ is free to take whatever value each age requires. And the distinction between geometries lives in the tail. Near the peak the curves are nearly indistinguishable, and they separate only beyond $J_z\simeq J_0$.

\begin{figure*}[t]
\centering
\includegraphics[width=\textwidth]{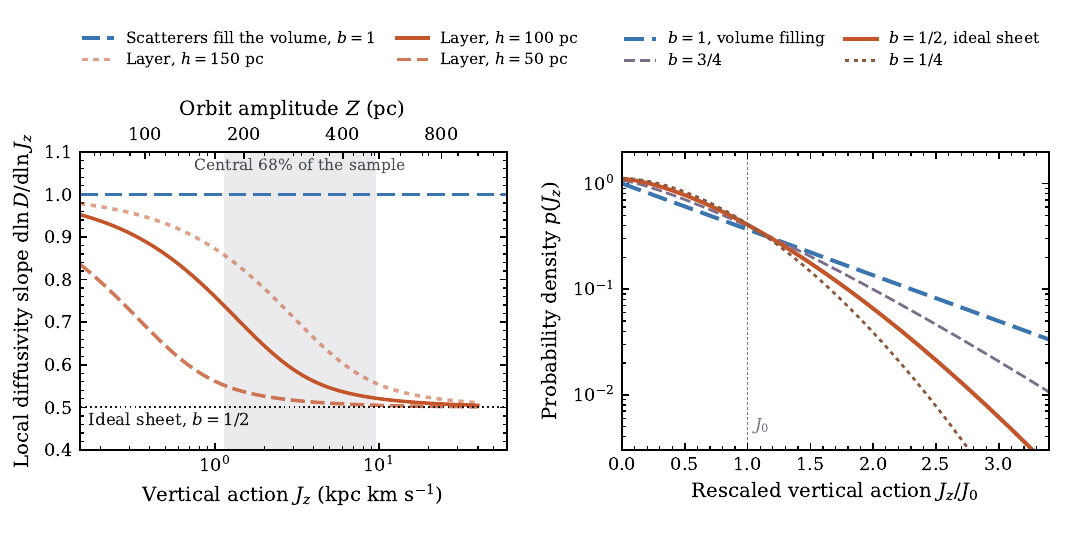}
\caption{From the scattering profile to the observable shape. The left panel shows the local diffusivity slope $b=\dd\ln D/\dd\ln J_z$ that a Gaussian scattering profile of scale height $h$ produces. The curves are drawn at $50$, $100$ and $150$\,pc, bracketing the $80$--$145$\,pc computed from the cloud catalogue, and Table~\ref{tab:geometry} gives the catalogue-derived thicknesses and the exponents they predict. The upper axis converts the action into the height the orbit reaches, $Z=\sqrt{2J_z/\nu}$, with the $\nu=75\,\kms\,{\rm kpc}^{-1}$ used throughout. The shaded band spans the central $68$ per cent of the observed actions, whose orbits reach several times the thickness of the layer. The right panel shows the coeval distribution each geometry produces, with the scale $J_0$ marked. The exponent of the tail is $2-b$, and the geometries are indistinguishable near the peak and separate only beyond $J_0$.}
\label{fig:profiles}
\end{figure*}

\subsection{The exponent implied by each geometry}

The exponent has been carried as a free parameter throughout. Its value for a given arrangement of the scatterers follows from the single integral of equation~\eqref{eq:master}, evaluated for the two limiting cases. If the scatterers fill the volume, $q=q_0$ comes out of the integral and leaves the area of a semicircle, so
\begin{equation}
D=\frac{q_0}{\pi}\cdot\frac{\pi Z^{2}}{2}=\frac{q_0Z^{2}}{2}=\frac{q_0}{\nu}J_z ,
\qquad b=1 .
\label{eq:uniform}
\end{equation}
A star whose orbit reaches higher moves faster, remains in contact with the perturbers just as often, and its diffusivity grows in exact proportion to the action it already has.

If instead they are confined to a layer thin compared with the orbit, then wherever $q$ is non-zero $|z|\ll Z$ and the window is flat at $\sqrt{Z^2-z^2}\simeq Z$, so only the column $Q\equiv\int q\,\dd z$ survives,
\begin{equation}
D\simeq\frac{Z}{\pi}\int q\,\dd z=\frac{QZ}{\pi}
=\frac{Q}{\pi}\sqrt{\frac{2J_z}{\nu}} ,
\qquad b=\tfrac12 .
\label{eq:sheet}
\end{equation}
A star whose orbit reaches higher is hit harder, because it crosses the plane faster and the action change goes as $v_z$. At the plane the energy is all kinetic, so $v_z^{2}(0)=2E_z=2\nu J_z$. It is hit for less time, because the dwell time inside a layer of fixed thickness goes as $1/v_z(0)\propto J_z^{-1/2}$. Multiplying,
\begin{equation}
D\propto J_z\times J_z^{-1/2}=J_z^{1/2} .
\end{equation}
One of the two powers of the crossing speed is cancelled by the brevity of the crossing, the growth of heating with action is halved, and the tail exponent rises from $1$ to $3/2$.

The competition set out qualitatively in Section~\ref{sec:physical} is the difference between $b=1$ and $b=1/2$. Where the scatterers fill the volume the star is among them at every point of its orbit, so only the strength of the impulses counts and the diffusivity follows the action itself. Where they lie in a layer the star meets them only at its crossings, so the brevity of those crossings counts as well, and it removes half of the growth. Figure~\ref{fig:profiles} shows both, together with what a layer of finite thickness gives in between and the distributions that follow.

Neither the number of perturbers, nor their masses, nor the strength of the kicks entered either result. Those quantities set $q_0$ and $Q$, which fix the amplitude $D_0$ and therefore how fast the distribution broadens. They do not touch the exponent. That separation is what allows the geometry to be measured without knowing the heating history.

Equation~\eqref{eq:solution} and the two geometries together are the result the rest of the paper uses. The exponent of the tail is $2-b$, so the geometries are distinguished by how sharply the distribution bends, an exponential for scatterers filling the volume and a three-halves stretched exponential for a midplane layer.

Earlier work reached parts of this by separate routes. The cases $b=1$, $b=1/2$ and $b=0$ appear in different scattering models of \citet{binneylacey1988}, and \citet{jenkins1992} later added finite cloud-layer thickness as a correction to their coefficients. Appendix~\ref{app:similarity} sets out those routes, including the $b=1/2$ they reached in the radial action from a different origin, and Appendix~\ref{app:thickness} inserts a layer of finite thickness. Written with $Z^{2}$ as the coordinate, equation~\eqref{eq:master} is an Abel transform, so the map from the height profile of the scatterers to $D(J_z)$ runs in both directions.

\section{Measurement of the Diffusivity Slope}
\label{sec:measurement}

The theory leaves a one-parameter family, with the exponent set by the geometry and the scale free at every age. Asking the data which member they belong to means handling three things first, each of which can imitate a change in the exponent. A real sample mixes stars of many ages, the survey does not see them all equally, and the ages and actions carry errors.

\subsection{Data and sample selection}

We use the archived per-radius products of \citet{tingrix2019}, which supply vertical actions, ages and a star-by-star selection function for low-$\alpha$ red clump stars drawn from APOGEE DR14 \citep{majewski2017,tinghawkinsrix2018}, with proper motions from \textit{Gaia} DR2 \citep{gaia2018}. The distances are spectro-photometric rather than parallaxes, which is what makes the sample usable several kiloparsecs away. Red clump stars serve here because their luminosity function is narrow enough to make them standard candles \citep{paczynskistanek1998,alves2000,bovy2014}, and because asteroseismically calibrated ages are available for them.

Later releases hold more stars, and there are two reasons for using these products rather than rebuilding the sample from a newer one. The actions, the ages and the star-by-star selection function were derived and published together, so the measurement inherits a chain that has already been vetted end to end. The limiting precision is also not statistical. The exponent has to be compared against the prediction for a real molecular layer, and the finite thickness, the encounter speed and the anharmonic potential of Appendices~\ref{app:thickness} and~\ref{app:anharmonic} spread that prediction over $0.44$--$0.54$, a range comparable to the uncertainty on the measurement itself. A larger catalogue would sharpen the measurement without sharpening what it is measured against.

Every star needs a vertical action, an age, and a radius at which its heating happened. The vertical action of a star is the phase-space area of its vertical oscillation, and computing it requires a position, a velocity and a potential. Positions come from APOGEE sky coordinates together with spectro-photometric distances. The distances use \textit{WISE} $W1$ photometry, with extinction estimated from the \textit{Gaia} $G-W1$ colour, and are precise to $\sigma_d/d\simeq7$ per cent \citep{hawkins2017,tingrix2019}. Velocities combine APOGEE radial velocities with \textit{Gaia} proper motions, so the transverse velocity carries the distance error directly. The actions were then computed with \texttt{galpy} \citep{bovy2015} in its Milky Way potential, scaled to the adopted values $R_\odot=8.2$\,kpc and $v_c(R_\odot)=240\,\kms$ \citep{schonrich2012,blandhawthorngerhard2016}, and using a solar motion of $(U,V,W)=(11.1,\,12.24,\,7.25)\,\kms$ \citep{schonrich2010}.

These are full-potential actions rather than the harmonic ones of Section~\ref{sec:theory}. Actions are well defined without the harmonic approximation in any integrable potential \citep{binneytremaine2008}, so the data are not being forced into that approximation. The harmonic limit enters only in the step from $q(z)$ to $D(J_z)$, and Appendix~\ref{app:anharmonic} carries that step through a general potential to show what it costs. The fractional action error is roughly twice the fractional distance error, because $J_z$ grows as $v_z^{2}$ near the plane while $v_z$ inherits the distance error linearly. The distances alone therefore give about $14$ per cent, close to the $15$ per cent \citet{tingrix2019} quote, and we adopt $20$ per cent to leave room for the radial velocities, the proper motions and the potential.

Ages are spectroscopic, inferred from the APOGEE spectra by a neural network trained on APOKASC-2 asteroseismic ages \citep{pinsonneault2018}, and are precise to about $25$ per cent \citep{tingrix2019}. An age error carries into the shape, because the comparison is between a star's action and the scale appropriate to its age. Age enters the model through that scale and nowhere else, so an error in $\tau$ moves $J_0$ and with it the ratio the shape is fitted to, at a level not small against the action error itself. Both are folded into the likelihood.

The third quantity is the radius at which a star did its heating, which is neither its birth radius nor its present one, since stars migrate. Following \citet{tingrix2019} we use the effective radius $R_{\rm eff}$, the average of the current radius and a birth radius estimated from the age--metallicity relation and a calibrated migration model \citep{frankel2018}. It is a coarse proxy, but the exponent is measured at fixed radius with the scale free, so an error in $R_{\rm eff}$ blurs the radial gradient rather than biasing the shape. The radial actions used in Section~\ref{sec:amplitude} come from the same parent catalogue, matched star by star.

We keep stars with $5\le R_{\rm eff}\le10$\,kpc, inside which the distances are reliable and the disc is not yet warped, and with $\tau>2$\,Gyr, so that accumulated heating dominates the action a star was born with. That leaves $7589$ stars with a median effective radius of $7.9$\,kpc.

The solution of equation~\eqref{eq:solution} describes a coeval population a time $t$ after it was born cold, so for a star of age $\tau$ the diffusion clock has run for $t=\tau$. A catalogue holds many such populations side by side, one at each age, so a single snapshot of the sky supplies the whole time sequence without waiting for it.

\subsection{The selection function}

APOGEE observes along fixed lines of sight and is magnitude limited, so at a given distance it recovers stars near the plane more efficiently than stars whose orbits carry them high above it. The loss therefore falls on the tail, which is the part of the distribution where the two geometries separate. Left uncorrected it removes tail stars and drives the exponent upwards, towards the volume-filling value.

The effect is large and it varies from star to star, so it needs a per-star account rather than a global one. \citet{tingrix2019} computed the visible fraction on a grid in $(J_z,R_{\rm eff})$ by integrating representative circular-orbit vertical trajectories in the same potential, assumed no additional dependence on $J_R$, and interpolated that grid to each star. They summarised it as
\begin{equation}
S_i(J_z)=
\begin{cases}
1, & J_z<J_{\rm t}(R_{\rm eff}),\\[2pt]
c(R_{\rm eff})\,J_z^{-\xi(R_{\rm eff})}, & J_z\ge J_{\rm t}(R_{\rm eff}),
\end{cases}
\label{eq:selection}
\end{equation}
in which $R_{\rm eff}$ is the star's effective radius, $J_{\rm t}$ the transition action, $\xi$ the power-law index and $c$ its coefficient. Below the transition an orbit stays inside the slab for its whole period, and above it the visible fraction falls as a power of the action.

All three parameters are low-order polynomials in radius, which \citet{tingrix2019} fitted to the numerically integrated selection,
\begin{equation}
\begin{aligned}
J_{\rm t}&=-0.013205\,R_{\rm eff}^{3}+0.50325\,R_{\rm eff}^{2}\\
&\quad-6.7793\,R_{\rm eff}+36.481 ,\\[2pt]
\xi&=-5.835\times10^{-5}\,R_{\rm eff}^{3}+0.0029564\,R_{\rm eff}^{2}\\
&\quad-0.056283\,R_{\rm eff}+1.0056 ,\\[2pt]
c&=0.0013318\,R_{\rm eff}^{4}-0.060855\,R_{\rm eff}^{3}+1.0425\,R_{\rm eff}^{2}\\
&\quad-8.1934\,R_{\rm eff}+27.776 ,
\end{aligned}
\label{eq:selpoly}
\end{equation}
with $R_{\rm eff}$ in kpc and $J_{\rm t}$ in $\kpckms$, so each star's radius sets its own selection. The radial dependence is there because the transition marks a height rather than an action. Reaching a height $z$ costs an action $\nu z^{2}/2$, so the weaker restoring force at larger radius puts the top of the slab at a smaller action, and the transition falls from about $12$ to $6\,\kpckms$ between $5.5$ and $9.5$\,kpc, the factor of two in $\nu$ that the declining surface density implies. The median index is $\xi=0.72$ and the median transition action is $7.9\,\kpckms$, close to the scale $J_0$ itself, and across the sample $21$ per cent of the stars sit above their own transition. Stars with $J_z>20\,\kpckms$ are recovered with a median efficiency of $33$ per cent. Efficiency falls with action and, more weakly, with radius, so the worst case in the sample is a single high-action star near the outer edge, whose efficiency is $16$ per cent.

We apply the selection to the model rather than to the data, multiplying by $S_i$ and renormalising. Dividing the data by $S_i$ instead reweights individual stars, so the few tail stars that survive acquire large weights and the noise in the tail is amplified. Multiplying leaves the data untouched and keeps the likelihood a proper probability for each star.

\begin{figure*}[t]
\centering
\includegraphics[width=\textwidth]{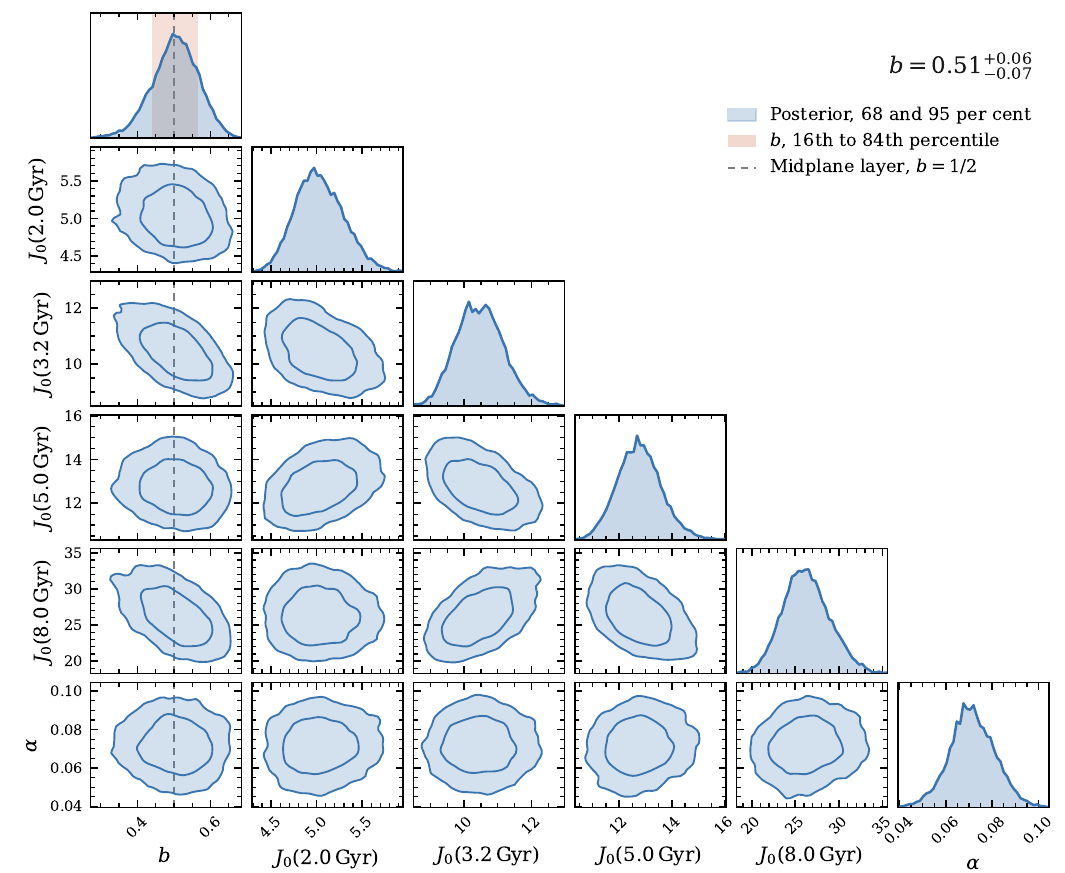}
\caption{The fit to the $7589$ stars between $5$ and $10$\,kpc and $2$ and $8$\,Gyr old. Contours enclose $68$ and $95$ per cent of the posterior samples over the six free parameters. These are the diffusivity slope $b$, which sets the shape of the distribution through $p(J_z)\propto\exp[-(J_z/J_0)^{2-b}]$, the scale $J_0$ at the four age knots $2.0$, $3.2$, $5.0$ and $8.0$\,Gyr, in $\kpckms$, and the radial log-slope $\alpha$ in ${\rm kpc}^{-1}$, by which $\log J_0$ tilts with radius about $8$\,kpc. They carry the heating history, left free so that the exponent is measured without assuming any form for it. The shaded band spans the $16$th to $84$th percentiles of the marginal for $b$. The dashed line marks the midplane value $b=1/2$. The exponent is uncorrelated with $\alpha$ and anti-correlated with the four knot scales $J_0(\tau_k)$.}
\label{fig:posterior}
\end{figure*}

\subsection{The likelihood}

The whole heating history enters equation~\eqref{eq:solution} through $J_0$ alone, so leaving $J_0$ free as a function of age removes the history from the problem without any assumption about what it was. The same freedom covers a second difficulty. The solution was derived for a population born cold, and stars are not born cold. They inherit the vertical motion of the gas they form from, which changes with epoch and with radius. But the family of equation~\eqref{eq:solution} is closed under its own heating. A population born with the same shape and a scale $J_{\rm b}$ is described, after a time $\tau$, by the same expression with $J_0^{\,2-b}$ replaced by $J_{\rm b}^{\,2-b}+\tfrac12(2-b)^{2}\int_0^{\tau}D_0\,\dd t$. A birth distribution therefore shifts the scale and leaves the shape alone. Fitting $J_0$ freely absorbs the birth term for the same reason it absorbs the history, and the exponent is untouched by either. That holds exactly when the birth population has the shape of the family, and Appendix~\ref{app:validation} injects one that does not.

We carry $J_0$ as four free values $\ell_k\equiv\log J_0(\tau_k)$ at knots equally spaced in $\log\tau$ between the youngest and the oldest observed age, together with one log-slope in radius,
\begin{equation}
\log J_0(\tau,R)=L(\log\tau)+\alpha\,(R-8\,{\rm kpc}) ,
\label{eq:knots}
\end{equation}
with $L$ linear between consecutive knots and constant outside them,
\begin{equation}
L(\log\tau)=
\begin{cases}
\ell_1 , & \tau\le\tau_1 ,\\[4pt]
\ell_k+\frac{\big(\ell_{k+1}-\ell_k\big)\log(\tau/\tau_k)}
{\log(\tau_{k+1}/\tau_k)} ,
& \tau_k\le\tau\le\tau_{k+1} ,\\[6pt]
\ell_4 , & \tau\ge\tau_4 ,
\end{cases}
\label{eq:knotshape}
\end{equation}
which is four numbers and a slope. The radial term is there because both the gas surface density and the restoring frequency change across the annulus, so the amplitude of the heating should not be forced to be the same at $5$ and $10$\,kpc. No form is assumed for the growth of $J_0$ with age, and a single $b$ is shared by every star at every age and radius.

A star's age, action and radius are each measured with an uncertainty, so its latent true values are unknown. The likelihood must average the model over those latent values, weighted by how likely each is given the measurement. Drawing random samples would leave the likelihood noisy and hard to optimise, so each average is done by Gaussian quadrature instead. Each error is lognormal, so in the logarithm of the variable it becomes a Gaussian weight on the whole line. Gaussian quadrature replaces such an integral by a short weighted sum over nodes $k$, carrying weights $w_{ik}$ for star $i$, with the nodes placed so the sum is exact for polynomials up to degree $2n-1$ with $n$ of them, which is why four points do the work of a fine grid. For a Gaussian weight those are the Gauss--Hermite nodes, so one rule serves age, radius and action alike. Some of the age nodes fall outside the range the knots span, which is why $L$ is held constant beyond them. They carry $10.7$ per cent of the total weight below the youngest knot and $2.0$ per cent above the oldest, so what is assumed there is not negligible.

Each star therefore carries a small grid of trial values rather than a single measured one. Collecting the pieces, its likelihood $\mathcal{L}_i$ is
\begin{equation}
\mathcal{L}_i=\sum_k w_{ik}\int_0^{\infty}\!\dd J_z\;
p(J_i\mid J_z)\;\frac{p(J_z\mid J_{0,ik})}{Z_{ik}}\;S_i(J_z) ,
\label{eq:likelihood}
\end{equation}
in which $p(J_i\mid J_z)$ is the action error kernel carrying a true action to the observed $J_i$, $p(J_z\mid J_{0,ik})$ is the solution of equation~\eqref{eq:solution} at the scale that node implies, and $S_i(J_z)$ is the selection function of equation~\eqref{eq:selection}. The normalisation
\begin{equation}
Z_{ik}=\int_0^{J_{\max}}\!\dd J_z'\;p(J_z'\mid J_{0,ik})\,S_i(J_z')
\label{eq:norm}
\end{equation}
keeps each component a probability over the observable range, $k$ runs over the grid of latent ages and radii, and $J_{\max}$ bounds that range. The scale $J_{0,ik}$ is evaluated from equation~\eqref{eq:knots} at each node's own latent age, so populations whose true scales differ are never mixed. Each latent component is normalised before the components are mixed, which the conditional error kernel requires.

Equation~\eqref{eq:norm} is the one integral needed for every star at every trial parameter set, and it is analytic in $(J_0,b)$, a combination of incomplete gamma functions given in Appendix~\ref{app:likelihood}. No Monte Carlo integration enters the likelihood, so it is smooth in its parameters and cheap enough to sample. The posterior is taken by MCMC under flat priors.

\begin{figure}[t]
\centering
\includegraphics[width=\columnwidth]{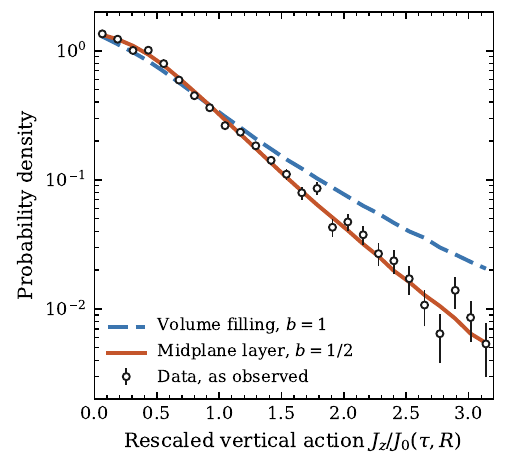}
\caption{The measured shape, from the same $7589$ stars. Each star's action is divided by the scale $J_0$ that the fit assigns to its age and radius, which by equation~\eqref{eq:solution} puts every age on one curve. The vertical axis is probability density, logarithmic, against that rescaled action. The points are the observed stars, and the two curves are the idealised geometries $b=1/2$ and $b=1$, each drawn as a synthetic population passed through the same selection function, errors and rescaling as the data. The midplane shape follows the data over the full range, whereas the exponential departs from it beyond $J_z\simeq J_0$.}
\label{fig:universal}
\end{figure}

\begin{figure*}[t]
\centering
\includegraphics[width=\textwidth]{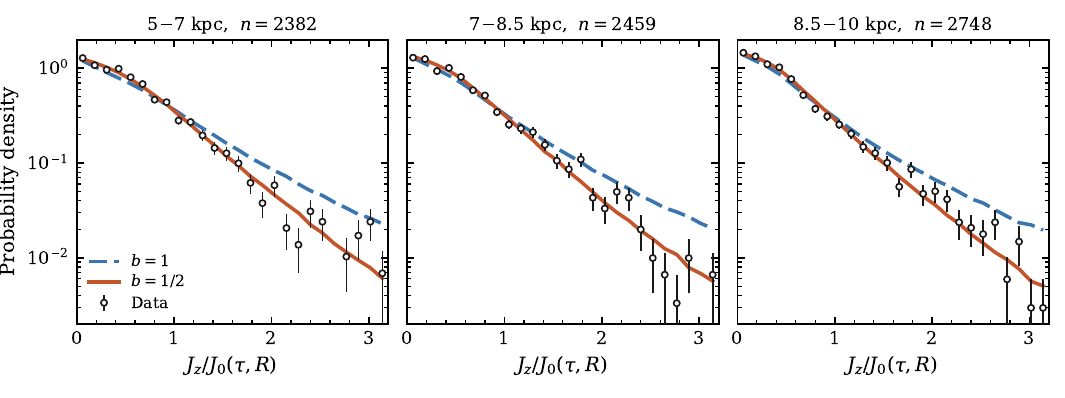}
\caption{The same comparison as Figure~\ref{fig:universal}, made separately in three annuli in $R_{\rm eff}$, $5$--$7$, $7$--$8.5$ and $8.5$--$10$\,kpc, cut to hold about a third of the stars each. The scale $J_0$ is the one the global fit assigns, knots and radial slope together, so no panel is refitted. The tail is correspondingly noisier than in the stack. The data follow the midplane curve and fall below the volume-filling one beyond $J_z\simeq J_0$ in all three.}
\label{fig:radial}
\end{figure*}

\subsection{The measured exponent}
\label{sec:exponentresult}

Fitted to the $7589$ stars, with the four knot scales and the radial slope free as well, the posterior over all six parameters is the one Figure~\ref{fig:posterior} shows. The exponent is $b=0.51^{+0.06}_{-0.07}$, the median of its marginal and the $16$th and $84$th percentiles. It is uncorrelated with the radial slope, so the radial gradient and the shape are measured independently of one another. It is anti-correlated with the four knot scales $J_0(\tau_k)$, which is the sign to expect, since raising $b$ lowers the tail exponent $2-b$ and flattens the tail, and the fit offsets that by shrinking the scale. The anti-correlation is not spread evenly across the knots but alternates between neighbours, because the scale history is piecewise linear and a shift at one knot is partly undone by the two beside it.

Against the two idealised geometries, this is the midplane result. The scatterers act where the star crosses the plane, and not throughout the volume its orbit sweeps. Injecting each geometry into synthetic catalogues built on the same stars, cuts, selection function and error model recovers both unbiased, separates them by $5.6\sigma$, and returns a known exponent with a scatter of $0.078$, as Appendix~\ref{app:validation} sets out.

The fit assigns each star a scale $J_0$ from its own age and radius, and equation~\eqref{eq:solution} makes the ratio $x=J_z/J_0$ follow $\exp(-x^{2-b})$ whatever the age. Forming that ratio collapses every age onto a single curve, which is the shape being measured. Two things keep that curve from being equation~\eqref{eq:solution} itself. The selection function of equation~\eqref{eq:selection} has already thinned the tail, which is the part the comparison rests on. And the rescaling can only use each star's observed age, while its action was set by the latent true age the likelihood integrates over, so the mismatch scatters $x$ even where the model is exact. The curve drawn for each geometry is therefore a synthetic population at that $b$, put through the same selection function, the same errors and the same rescaling as the stars.

Figure~\ref{fig:universal} shows the comparison. The midplane shape follows the data across the whole range. The exponential lies above the data from $x\simeq1$ outwards. Splitting the stars into three annuli of nearly equal size leaves the same picture in each, as Figure~\ref{fig:radial} shows, so the shape is not an average over annuli that differ. Each annulus holds only a third of the stars, so the stack is where the separation is clearest.

Both geometries are idealisations, and three corrections separate them from the sample. A real molecular layer has a finite thickness, the encounter speed is not independent of the action, and the vertical potential is not harmonic. Appendices~\ref{app:thickness} and~\ref{app:anharmonic} evaluate all three.

\begin{table}[!tb]
\centering
\caption{The exponent across the choices of sample and assumed error, each a separate fit of the likelihood of Section~\ref{sec:measurement}. The columns are the radial range in $R_{\rm eff}$, the youngest age kept, with $8$\,Gyr the oldest throughout, the fractional action error assumed, the number of stars, and the exponent. The first row is the configuration adopted, and gives the posterior median. The other rows are maximum-likelihood fits. The last two rows are what each geometry predicts for this sample, as a range because the vertical potential is not harmonic, and Table~\ref{tab:geometry} sets out how they are built.}
\label{tab:configurations}
\small
\hspace*{-41pt}\begin{tabular}{ccccc}
\toprule
$R$ (kpc) & $\tau_{\min}$ (Gyr) & $\sigma_{J_z}/J_z$ & $n$ & $b$\\
\midrule
$5$--$10$ & $2$ & $0.20$ & $7589$ & $0.506$\\
$5$--$10$ & $2$ & $0.15$ & $7589$ & $0.568$\\
$6$--$10$ & $2$ & $0.20$ & $6606$ & $0.483$\\
$6$--$10$ & $2$ & $0.15$ & $6606$ & $0.553$\\
$5$--$10$ & $3$ & $0.20$ & $4620$ & $0.553$\\
$6$--$10$ & $3$ & $0.20$ & $3835$ & $0.554$\\
$7$--$9 $ & $3$ & $0.20$ & $1956$ & $0.506$\\
\midrule
\multicolumn{4}{c}{Midplane layer predicts} & $0.44$--$0.54$\\
\multicolumn{4}{c}{Volume filling predicts} & $0.85$--$1.00$\\
\bottomrule
\end{tabular}
\end{table}

Applying our mass and stellar-sample radial weights to the catalogue of \citet{miville2017} gives an rms spread of about $100$\,pc over $5$--$10$\,kpc, rising from roughly $80$\,pc in the inner annulus to $145$\,pc in the outer one as the layer flares. Thickness and encounter speed act in opposite directions, with the encounter speed the larger, and the anharmonic potential pushes back the other way. Localised scattering then predicts $0.44$--$0.54$ for the sample, which brackets the measurement. Volume-filling scatterers have no layer to thicken, so only the other two corrections apply to them, and they give $0.85$--$1.00$. The two bands do not overlap, and neither correction closes the gap, since both move the geometries the same way.

The result should not turn on where the sample boundaries are drawn. Table~\ref{tab:configurations} varies the radial range, the age cut and the assumed action error. The rows span $0.483$ to $0.568$, a half-range of $0.043$, so the sample definition adds no systematic beyond the noise of the estimator. The assumed action error matters most, and in the direction one expects, since under-stating it leaves error in the tail for the exponent to absorb.

\begin{figure*}[!t]
\centering
\includegraphics[width=\textwidth]{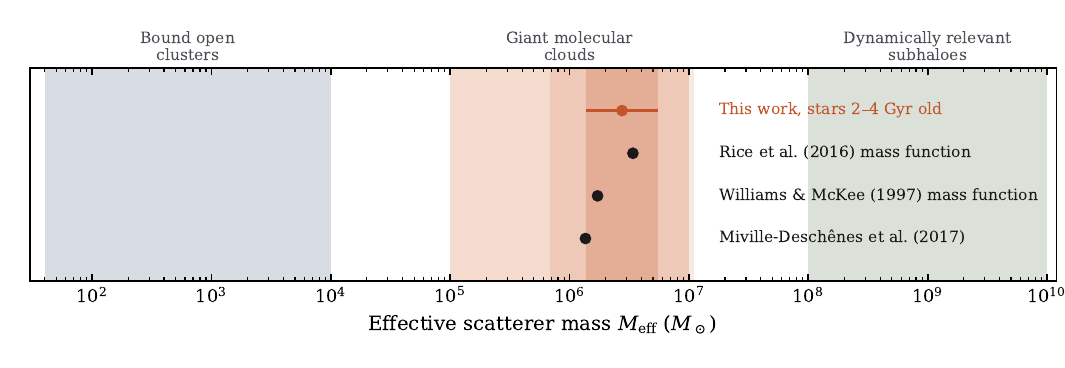}
\caption{The effective scatterer mass implied by the heating amplitude, compared with three values from the cloud side. The comparison bands mark the inferred bound-open-cluster range, the GMC scale, and the tidal-mass threshold $M_{\rm tid}\ge10^{8}\,\Msun$ above which \citet{moetazedianjust2016} select dark-matter subhaloes as potentially relevant vertical perturbers. Only the $2$--$4$\,Gyr stars are shown, since their answer averages over the shortest and most recent stretch of heating and is the one comparable to a catalogue of the clouds that are here now. Their statistical uncertainty is five per cent, far smaller than the systematic terms of Section~\ref{sec:discussion}, so the shading about the stellar value of $2.7\times10^{6}\,\Msun$ shows factors of two and four.}
\label{fig:amplitude}
\end{figure*}

The four knot scales rise from $J_0=5.0$ to $26.2\,\kpckms$ across the age range, that is $J_0\propto\tau^{1.12}$, where equation~\eqref{eq:clock} at $b=1/2$ with a constant rate would give $\tau^{0.67}$. The disc heated faster in the past than a fixed rate allows, which is the conclusion \citet{tingrix2019} reached from the growth of the mean action. The scales do not separate that from a disc whose older stars were born hotter, as Section~\ref{sec:limitations} sets out. A rate that varies with time is what the free scales absorb, so it changes $J_0$ at each age and leaves the shape the exponent is read from untouched.

\section{The Heating Amplitude and the Implied Scatterer Mass}
\label{sec:amplitude}

\begin{figure*}[!t]
\centering
\includegraphics[width=\textwidth]{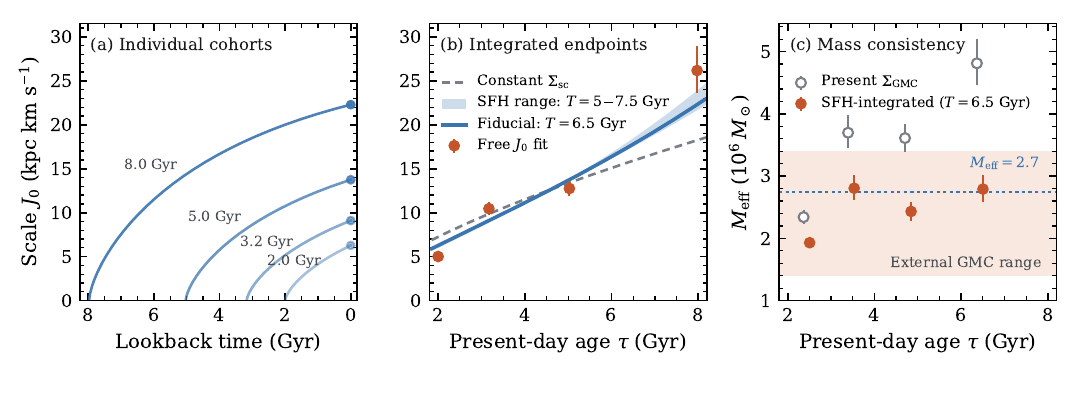}
\caption{The freely measured heating scales against an evolving molecular-scatterer density. The history is applied only after the free-scale fit, so it moves none of the knots and does not touch the exponent. The left panel follows the scale $J_0$ of four coeval populations from birth to the present under the fiducial exponential star-formation history, against lookback time, their endpoints tracing the curve in the middle panel. The middle panel plots the four knot scales $J_0(\tau_k)$ against present-day age, and compares them with a constant scatterer density and with exponential histories, each carrying one vertical normalisation. The solid line uses $T=6.5$\,Gyr and the band spans $5$--$7.5$\,Gyr. The right panel shows the effective masses on the same age axis, obtained by dividing each knot first by the present molecular surface density and then by its lifetime-integrated value under the fiducial history. The shaded interval is the $1.4$--$3.4\times10^{6}\,\Msun$ range from the cloud determinations of Figure~\ref{fig:amplitude}, and the dotted line marks the $2.7\times10^{6}\,\Msun$ estimate from the combined $2$--$4$\,Gyr sample.}
\label{fig:sfh}
\end{figure*}

The exponent came from the shape of the distribution at fixed age, and the free scales absorbed the strength of the heating so that it never entered the answer. Those scales are now the measurement. The exponent says how the scatterers are arranged. How fast those scales grow says how much mass they carry. A cloud of mass $M$ passing a star at impact parameter $\varpi$ with relative speed $V$ delivers a perpendicular impulse $2GM/(\varpi V)$ in the impulse approximation. Summing the mean square impulse over a population of clouds gives every decade of encounter distance equal weight, and produces the Coulomb logarithm $\ln\Lambda=\ln(\varpi_{\max}/\varpi_{\min})$ \citep{binneytremaine2008}. The mass integral leaves the second moment of the mass function.

Let $N(M)$ be the number of clouds per unit area of disc per unit mass. Writing $\Ssc\equiv\int MN\,\dd M$ for the scatterer surface density and $\Meff\equiv\int M^{2}N\,\dd M/\int MN\,\dd M$ for the mass-weighted mean mass, the column through the layer is
\begin{equation}
Q=\frac{4\pi G^{2}\,\Ssc\,\Meff\,\ln\Lambda}{V} .
\label{eq:Q}
\end{equation}
Appendix~\ref{app:impulse} carries the sum through. Which moment of the mass function the heating responds to is fixed by the derivation, because the impulse is linear in $M$ while the diffusion is quadratic in the impulse. Two populations of the same total mass therefore heat at different rates if their mass functions differ. For an observed mass function $\dd N/\dd M\propto M^{-1.7}$, the integral $\int M^{2}N\,\dd M$ is controlled by the upper cutoff, so $\Meff$ is a statement about the largest clouds.

The cloud layer is about $100$\,pc thick, against orbits that reach $200$ to $600$\,pc across the middle two thirds of the selection-corrected sample, so the thin-layer form applies. Putting that column into the thin-layer diffusivity and running the clock at $b=1/2$ turns it into a prediction for how fast a coeval population broadens,
\begin{equation}
J_0^{3/2}(t)=\frac{9\,G^{2}\,\Ssc\,\Meff\,\ln\Lambda}{2\,V}
\sqrt{\frac{2}{\nu}}\;t .
\label{eq:amplitude}
\end{equation}

Every term in equation~\eqref{eq:amplitude} but $\Ssc\Meff$ is either measured per star or held fixed. For each star its own age fixes $t$, and its own orbit fixes the encounter speed. Clouds move slowly against the stars, so $V$ is close to the star's own random velocity, and $V^{2}\simeq2E_R+2E_z$ splits into an in-plane and a vertical part. The in-plane part is the epicyclic energy $E_R=\kappa\JR$, with $\kappa$ the epicyclic frequency, and the radial actions come from the parent catalogue, whose match to the sample is exact and one-to-one. We take $V\simeq\sqrt{2\kappa\JR}$ and drop the vertical part, which is the smaller of the two for four fifths of the stars. The vertical frequency is held at $\nu=75\,\kms\,{\rm kpc}^{-1}$ and the Coulomb logarithm at $\ln\Lambda=3$, which Appendix~\ref{app:impulse} brackets between $2.8$ and $4.7$ from the cloud sizes in the catalogue. That leaves $\Ssc\Meff$.

Section~\ref{sec:measurement} measured $J_0$ at each age with the scale left free, so equation~\eqref{eq:amplitude} can be read backwards on those scales rather than run forwards from an assumed rate. Each returns the amplitude that produced it, one value per star.

The $2$--$4$\,Gyr stars give the cleanest comparison with a present-day cloud catalogue, since they average over the shortest stretch of history. Over this combined sample,
\begin{equation}
\Ssc\Meff=8.8\times10^{6}\,\Msun^{2}\,{\rm pc}^{-2} ,
\end{equation}
with a statistical uncertainty of five per cent inherited from the scales. The inversion is done at $b=1/2$, and redoing it at the measured $0.51$ moves the answer by under one per cent.

Turning $\Ssc\Meff$ into a mass needs a surface density, and that is where cloud data first enter. We take it from the catalogue of \citet{miville2017}, averaging over radius with the same weights the stellar sample carries, since the molecular surface density falls by a factor of seven across the sample. That gives $\SGMC=3.2\,\Msun\,{\rm pc}^{-2}$, and setting $\Ssc=\SGMC$,
\begin{equation}
\Meff\big|_\star=\frac{\Ssc\Meff}{\SGMC}=2.7\times10^{6}\,\Msun .
\end{equation}

Three values from the cloud side meet it, as Figure~\ref{fig:amplitude} shows. Summing the same catalogue cloud by cloud gives $1.4\times10^{6}\,\Msun$. The inner-Galaxy mass functions of \citet{williamsmckee1997} and \citet{rice2016}, evaluated with their stated lower limits, give $1.7$ and $3.4\times10^{6}\,\Msun$. The stellar value lies inside this factor-of-$2.5$ range, and all four agree on the scale, a few million solar masses. That is the GMC scale, above bound open clusters, which run from tens of solar masses to $10^{4}\,\Msun$ \citep{huntreffert2024}, and below the dark-matter subhaloes of tidal mass $M_{\rm tid}\ge10^{8}\,\Msun$ that \citet{moetazedianjust2016} select as potentially relevant vertical perturbers.

Splitting the stars into four age windows carries the same comparison further back. These are windows of stars, not the knots of the scale history, which stay where Section~\ref{sec:measurement} put them. Reading equation~\eqref{eq:amplitude} backwards within each window and dividing by today's molecular surface density gives $\Meff=2.3$, $3.7$, $3.6$ and $4.8\times10^{6}\,\Msun$ at median ages of $2.4$, $3.5$, $4.8$ and $6.4$\,Gyr, an apparent factor-of-two rise across the sample. Each bin, however, measures a lifetime average of $\Ssc\Meff/V$, so using today's $\SGMC$ for every bin assigns the entire history to $\Meff$. The scatterers themselves changed over that span, in abundance and in mass spectrum both.

How far the abundance alone carries the rise follows from the star-formation history, which is measured elsewhere and enters nothing that has been fitted here. Identify $\Ssc$ with the molecular surface density, and let the star-formation-rate surface density track it through the molecular Kennicutt--Schmidt relation $\Sigma_{\rm SFR}\propto\Ssc^{\NKS}$. For a star-formation history that declines exponentially forward in time with e-folding $T$, the density at lookback time $t_{\rm lb}$ is
\begin{equation}
\frac{\Ssc(t_{\rm lb})}{\Ssc(0)}
 = \left[\frac{\Sigma_{\rm SFR}(t_{\rm lb})}
 {\Sigma_{\rm SFR}(0)}\right]^{1/\NKS}
 = \exp\!\left(\frac{t_{\rm lb}}{\NKS T}\right) .
\label{eq:sfh-density}
\end{equation}
We take $\NKS=1$ and $T=6.5$\,Gyr, both following \citet{tingrix2019}. \citet{leroy2013} measure $\NKS=1\pm0.15$ across nearby discs as a first-order description with environmental variation around it, and $T$ sits inside the $6\pm1$\,Gyr of \citet{frankel2018} and their $5$--$7.5$\,Gyr selection variants. The time a cohort of age $\tau$ accumulates is then
\begin{equation}
\begin{aligned}
t_{\rm eff}(\tau)
&\equiv \int_{0}^{\tau}
 \frac{\Ssc(t_{\rm lb})}{\Ssc(0)}\,\dd t_{\rm lb}\\
&=\NKS T\!\left[\exp\!\left(\frac{\tau}{\NKS T}\right)-1\right]\\
&\xrightarrow{\NKS=1}T\!\left[\exp\!\left(\frac{\tau}{T}\right)-1\right] .
\end{aligned}
\label{eq:sfh-clock}
\end{equation}
The quantity $t_{\rm eff}$ is the duration of constant present-day-density heating that would produce the same accumulated signal, so the clock $t=\tau$ in equation~\eqref{eq:amplitude} becomes $t_{\rm eff}(\tau)$ with $\Meff$, $V$ and $\nu$ held fixed. The clock runs on each star's own age, as it does in equation~\eqref{eq:amplitude}. At the median age of the oldest bin, $6.4$\,Gyr, it gives $t_{\rm eff}=11.0$\,Gyr, so those stars were heated as though the present density had acted for $1.71$ times their age, while the youngest bin gains only a factor of $1.21$.

Those two factors are what the age trend needs. The four bins give $\Meff=1.9$, $2.8$, $2.4$ and $2.8\times10^{6}\,\Msun$, all inside the range the clouds give, and the oldest has fallen by $1.72$ against the youngest's $1.21$. A declining scatterer density therefore absorbs the apparent factor-of-two rise almost exactly, without moving a single measured scale. What is left over is scatter of a few tens of per cent, which four correlated knots cannot resolve into evolution of $\Meff$, of the encounter speed, or of the action at birth.

\section{Discussion}
\label{sec:discussion}

The two results read different things off the same fit. The exponent $b=0.51$ places the scatterers in a midplane layer rather than through the volume the orbit sweeps. The growth of the free scales fixes the heating amplitude. Comparing the combined $2$--$4$\,Gyr sample with the present molecular surface density turns it into $\Meff=2.7\times10^{6}\,\Msun$. Each can be set against an existing determination, the mass against what cloud catalogues and mass functions give and the exponent against the classical value of unity.

\subsection{The mass determinations compared}
\label{sec:massescompared}

The direct sum over the cloud catalogue and the two published mass functions of Section~\ref{sec:amplitude} span a factor of $2.5$ among themselves, and the stellar value falls inside that spread, a factor of two above the direct sum and just below the higher of the two. Being carried by the largest clouds, $\Meff$ depends on where a mass function is truncated and on how a survey divides blended emission into objects. The spread follows from those choices.

That spread understates the uncertainty, because every cloud number rests on the same CO-to-$\rm H_2$ factor $X_{\rm CO}=2\times10^{20}\,{\rm cm^{-2}\,(K\,km\,s^{-1})^{-1}}$, the value \citet{bolatto2013} recommend for the Milky Way disc with a $30$ per cent uncertainty, and it enters the comparison twice. The heating measures $\Ssc\Meff$, which uses no CO. Dividing it by a surface density proportional to $X_{\rm CO}$ makes our $\Meff$ scale as $X_{\rm CO}^{-1}$, while the catalogue's own effective mass scales as $X_{\rm CO}$, so the ratio between them goes as $X_{\rm CO}^{-2}$ and $30$ per cent becomes a factor of $1.7$.

The same catalogue supplies the gas mass we divide by, and that can be tested directly. Summing its clouds inside the annulus gives $7.9\times10^{8}\,\Msun$, which is $48$ per cent of its Galactic total. If star formation follows the molecular gas, that share of the Galactic rate of \citet{licquia2015} consumes the annulus in $1.0$\,Gyr. A depletion time means little on its own, so the benchmark has to come from other discs, and nearby ones give $2.2$\,Gyr with $0.3$\,dex of scatter \citep{leroy2013}, about one standard deviation away. The CO mass we assign to the scatterers is therefore consistent with the gas that makes stars.

What remains is a comparison across time. Each stellar value averages $\Ssc\Meff/V$ over the lives of the stars that supply it, whereas a catalogue counts the clouds that are here now, so any window reaching back into a gas-richer disc overstates $\Meff$. Section~\ref{sec:amplitude} shows the size of that, a factor of $1.71$ for the oldest bin and $1.21$ for the youngest, and the star-formation history of the low-$\alpha$ disc accounts for it. What the bins constrain is the product, so a larger $\Meff$ at earlier times would move the oldest points the same way as a faster-declining density or a colder disc. Two smaller terms also inflate the apparent mass. Stars are not born with zero action, so part of the observed heating was there from the start, and $\Meff$ scales inversely with $\ln\Lambda$ by equation~\eqref{eq:Q}, so the adopted $3$, near the bottom of the $2.8$ to $4.7$ that Appendix~\ref{app:impulse} brackets, carries tens of per cent of its own.

Not every term pushes that way. The encounter speed keeps only the in-plane motion, and restoring the vertical part would raise $\Meff$ by $38$ per cent. The largest term cannot be quantified. Vertical heating that in fact arrived from the in-plane direction is charged here to the clouds, so the amplitude bounds what the clouds supply rather than isolating it.

A stellar determination running above a cloud one is not new. \citet{jenkins1992} already found that the cloud heating parameter required to fit the observed age--velocity relation was far above what CO observations gave for its present value, and \citet{lacey1984} found that cloud masses had to sit near the top of the permitted range. Both set a lifetime-averaged heating against present-day cloud properties. The age-resolved scales show what that costs, and a heating determination reaching back several gigayears belongs against a gas history rather than against a catalogue of the present.

\subsection{The exponent inferred from local velocity distributions}
\label{sec:ageblind}

The same exponent was inferred from local velocity distributions once before, indirectly and with the opposite result. \citet{binneylacey1988} asked which diffusion coefficients are consistent with a velocity distribution that is Gaussian and stays Gaussian, and found that the vertical component of the action-space diffusion tensor must grow in proportion to the vertical energy, $D_{zz}\propto E_z$. Since $E_z=\nu J_z$, that is $b=1$. Two observations went into that, the vertical velocity distribution of nearby stars being Gaussian and the dispersion of a coeval population growing as $\sqrt{t}$. Both point to $b=1$, by different routes. The Gaussian shape does so directly. The growth does so through the clock of equation~\eqref{eq:clock}, which at a fixed heating rate gives $J_0\propto\tau^{1/(2-b)}$. In the harmonic limit $\sigma_z^{2}=\nu\langle J_z\rangle$, and $\langle J_z\rangle$ is proportional to $J_0$ at fixed shape, so a dispersion growing as $\sqrt{t}$ is a scale growing as $\tau^{1.0}$, which is $b=1$ exactly.

The shape fails on the sample rather than the physics. A sample selected without ages superposes populations of every age, each with its own scale $J_0(\tau)$. A mixture of scales has a heavier tail than any one of its members, and a heavier tail is read as a shallower fall-off. Fitting a single member of equation~\eqref{eq:solution} to such a mixture, in the limit of infinite sample size, returns an effective exponent rather than $b$. For a population that truly has $b=1/2$, formed at a constant rate over $0.1$--$10\,$Gyr and heated on the clock of equation~\eqref{eq:clock}, that exponent is $1.03$, which Appendix~\ref{app:blind} derives and Figure~\ref{fig:ageblind} draws. \citet{jenkins1992} reached the same conclusion independently, by Monte Carlo integration of the orbit-averaged equation, finding a continuously forming population close to isothermal while a coeval one develops significant sub-isothermal wings. How far the exponent moves depends on the star-formation history through the range of scales being mixed, so it is not a fixed offset that could be corrected for.

The growth fails because it constrains the geometry only when the rate is fixed and stars are born cold. Read that way our own free scales say much the same as theirs, rising as $\tau^{1.12}$ and so giving $b=1.10$. They are left free at each age precisely so that neither assumption is needed, and Section~\ref{sec:exponentresult} finds scales that no fixed rate with a cold start produces. The growth then measures that history and the geometry together, and the exponent has to come from the shape at fixed age.

The distribution itself was approached from several sides without the geometry being read off it. \citet{fujimoto1980} derived the position and velocity distributions of stars under short random cloud forces, and \citet{kokuboida1992} followed the three-dimensional velocity moments. Closest of all, \citet{villumsenbinney1985} fitted the endpoint of a coeval cloud-heating simulation with exponentials in both actions. An exponential is the $b=1$ form, but the two geometries differ only beyond $J_0$, and a fit to the body of a simulated distribution is insensitive to the tail that separates them. None of these attached the shape to the height profile of the scatterers.

The samples of that era could not have supplied the tail either, whether sorted by F-star ages \citep{carlberg1985} or by colour bins \citep{dehnenbinney1998}, neither of which yields a coeval distribution in vertical action. The exponential was also a standard ingredient of equilibrium disc models \citep{binney2010}, adopted as a fitting form rather than measured, so its success there is no evidence that heating produces it.

\subsection{The division of labour between clouds and spirals}
\label{sec:division}

The exponent also bears on a division of labour proposed long before these data. \citet{binneylacey1988} found that star--cloud scattering could not supply the radial diffusion coefficient the local kinematics seemed to require, and that the cloud sizes needed to repair it, of order a kiloparsec, would be sheared apart by Galactic rotation. That is a statement about the radial direction, and it leaves the clouds short of the in-plane energy budget. The measurement here is about the vertical direction, and it puts the scatterers in a thin layer at the midplane, which is where the clouds are. They constrain different directions and do not conflict.

Together they point at the picture \citet{jenkinsbinney1990} reached by a different argument, in which transient spiral structure supplies most of the in-plane energy and clouds convert part of it into vertical motion. \citet{barbaniswoltjer1967} had already set two conditions on large-scale structure if it is to account for the growth of the dispersion, that the perturbations be transient and that the disc heated more strongly in the past. The free scales of Section~\ref{sec:exponentresult} rise faster than a fixed rate allows, as a disc that heated more strongly in the past would.

The geometries make the division natural. Spiral structure is a distortion of the disc itself and is present wherever the star is, so a star feels it along the whole of its orbit. Molecular clouds sit in a layer thin compared with the vertical excursion of all but the youngest stars, so they act only during crossings, and the crossing is where an in-plane velocity is most efficiently turned into a vertical one. Simulations show the same asymmetry directly, with cloud heating slowing as stars leave the plane \citep{fujimoto2023} and gaseous structure changing the vertical action preferentially over the radial one \citep{arunima2025}. The exponent separates the mechanisms by where they act rather than by how much they heat, and it comes out at the thin-layer value.

The exponent settles which agent does the vertical scattering, not where the energy comes from. Diffusion in vertical action is confined near the midplane, which favours molecular clouds or any comparably localised structure in the cold interstellar medium. Anything the star feels throughout its orbit is ruled out as the main agent, whether dark substructure or the spiral distortion acting directly on the vertical motion. Spiral heating as a whole is not excluded by that. Energy a spiral puts into in-plane motion and a cloud crossing later turns upward carries the thin-layer signature, because the crossing is where the redirection happens. The shape counts that energy with the clouds. Only a perturbation felt throughout the orbit produces the volume-filling shape, and that is the channel the exponent limits.

The identification with clouds rests on less than the geometry does. Any population of independent scatterers confined to a thin layer gives the same shape, and only the amplitude of Section~\ref{sec:amplitude} names them. The amplitude in turn charges to the clouds whatever reached the vertical direction from in-plane motion, whichever agent redirected it, so the effective mass bounds them from above rather than counting them. Neither result says how much of the in-plane energy the spirals supply.

\subsection{Bounds on a second heating channel}
\label{sec:mixture}

The measurement returns one exponent, as a single scattering geometry must. If two channels act at once, the observed distribution is a mixture, and the question is how large the second component can be before the data notice.

We fit two channels together, one held at $b=1$ and one with $b$ free, requiring them to share a mean action at each age. Tying the means makes the test identifiable. Allowing them to float separately would let an exponential hide under the localised component at any fraction. The mixture is formed on normalised intrinsic densities before the selection function is applied, so the fraction refers to the underlying population rather than the observed one. The scale history remains free exactly as in the main fit. At zero fraction the model is the single-shape fit rewritten in the mean action, and returns the same likelihood and the same exponent.

Allowing the second channel improves the fit by $0.24$ in log likelihood, so the data do not ask for it. They give an upper limit instead, and how tight it is depends on what the localised channel is allowed to be. Left free in shape it lies below $49$ per cent at $95$ per cent confidence. The fit pays for the second channel by flattening the localised one, whose exponent falls out of the $0.44$--$0.54$ that a layer of the measured thickness predicts once the fraction passes $7$ per cent, reaching $0.31$ at the best fit and negative values by the bound. Requiring the localised channel to remain a layer therefore holds the volume-filling fraction below $7$ per cent. The likelihood is flat across that range, so the limit comes from the geometry rather than from the fit. A volume-filling component is therefore permitted a minority share of the diffusive heating and no more, whether it is supplied by bending waves, by dark substructure, or by anything else the star feels throughout its orbit. The shape measurement shows that the diffusive heating is midplane-localised, not that nothing else contributes.

The same bound settles a proposal the amplitude leaves open. The black-hole haloes of \citet{laceyostriker1985} weigh about $10^{6}\,\Msun$, within a factor of a few of what the amplitude returns here. Both hypotheses deliver the same $\Ssc\Meff$, so no measurement of a mass tells them apart. The exponent does, and without reference to any mass, because holes in a halo act on the star throughout its orbit and so give $b=1$.

\subsection{Relation to age--velocity dispersion measurements}

The quantity measured here is not the age--velocity dispersion relation. The closest comparison is with \citet{tingrix2019}, where the growth of the vertical action itself was fitted on the parent catalogue this sample is drawn from, giving a growth exponent near unity inside the solar radius. Our free scales give $J_0\propto\tau^{1.12}$ over $5$--$10$\,kpc, which is the same quantity measured on a subset of those stars, so the agreement is a consistency check rather than an independent one.

The routes need not agree, and different statistics are not the only reason. Multiplying the Fokker--Planck equation by $J_z^{\,2-b}$ and integrating gives an identity that holds for any birth distribution and any shape,
\begin{equation}
\big\langle J_z^{\,2-b}\big\rangle(\tau)
=\big\langle J_{\rm b}^{\,2-b}\big\rangle+\tfrac12(2-b)\,D_0\tau ,
\label{eq:additive}
\end{equation}
so the quantity that accumulates linearly in time is the $(2-b)$-th moment, not the second. In the harmonic limit $\sigma_z^{2}=\nu\langle J_z\rangle$, so the familiar fitting form $\sigma_z^{2}=\sigma_{\rm b}^{2}+C\tau^{2\beta}$ presumes that the first moment of the action is the additive one, which is true if and only if $b=1$. Fitting that form therefore builds in the geometry this paper is trying to measure, by a route independent of the age mixing of Section~\ref{sec:ageblind}.

Determinations that work in velocity dispersion rather than action do give a shallower growth, as that argument predicts. \citet{aumerbinney2009} find $\sigma_z\propto\tau^{0.45}$, \citet{sharma2021} $\sigma_z\propto\tau^{0.441\pm0.007}$ at fixed angular momentum, and \citet{sun2025} exponents that fall with radius, so no single number describes the disc \citep[see also][]{mackereth2019}. At fixed vertical frequency $\sigma_z\propto\tau^{0.44}$ corresponds to $J_z\propto\tau^{0.88}$, shallower than either action-based value. These are different statistics of different samples, but the direction of the difference is not accidental, and the measurement made here depends on none of these numbers.

The distinction is sharper than a difference of statistic. \citet{aumeravr2016} showed in $N$-body models that a measured age--velocity dispersion relation and the heating history of the stars that make it up have systematically different exponents, and that age errors lower the measured exponent further by smearing the ages. Both act on the relation between $\sigma$ and $\tau$, and neither acts on the shape at fixed age, which our model fits with the age errors marginalised over in the likelihood. They attribute the difference mainly to the declining importance of clouds relative to spiral structure and the bar over the life of the disc. Our scales rise as $\tau^{1.12}$ where a constant rate would give $\tau^{0.67}$, which is the stronger past heating that account requires, while the exponent keeps the vertical scattering midplane-localised throughout.

\subsection{Limitations of the diffusive description}
\label{sec:limitations}

The exponent is taken to be constant over the life of the disc, and it need not be. The layer that confines the scatterers was itself thicker when the gas fraction was higher, so early heating would have sat closer to the volume-filling case and drifted towards a sheet as the gas settled. That settling is the upside-down formation seen in cosmological simulations, in which the gas disc cools and thins with time \citep{bird2013,bird2021}. An exponent that drifts is not absorbed by the free scales the way a drifting amplitude is, because the scales enter the solution only through $J_0$ whereas the exponent sets its functional form. The measurement then returns an average over the ages in the sample, and measuring the trend would need a fit that let the exponent vary with age. Over the range fitted, though, the drift cannot be large. A spell near the volume-filling case early on would raise that average above the sheet value, and it comes out at $0.51$, with the volume-filling share held below $7$ per cent by Section~\ref{sec:mixture}. Whatever settling of this kind took place before $8$\,Gyr ago, between $5$ and $10$\,kpc it has left the geometry at the thin-layer value since. That the growth over these stars is accounted for by cloud heating is the conclusion \citet{tingrix2019} reached, and the exponent reaches it without the free scales having to separate a gas-richer past from a hotter birth.

The same settling makes the stars hotter at birth, and that bears on the free scales rather than on the exponent. \citet{bird2021} find a vertical dispersion of $6$--$8\,\kms$ for young stars in a simulation whose age--velocity relation matches the solar neighbourhood, with the birth dispersion rising towards earlier times. \citet{bird2013} argue that the trends of kinematics with age are largely imprinted at formation rather than by later heating. The scales absorb that amplitude, since $J_0(\tau)$ is free at every age whatever produced it. The climb in apparent $\Meff$ across the age bins is therefore not on its own evidence of a gas-richer past, because stars born hotter would produce the same trend, and the exponential star-formation history of Section~\ref{sec:amplitude} is one account of the climb rather than the only one.

A birth spread also changes the shape and cannot be absorbed completely by the free scales, so Appendix~\ref{app:validation} injects one. A birth action of $0.65\,\kpckms$, which is $\sigma_{\rm gas}^{2}/\nu$ for a $7\,\kms$ gas layer, moves the exponent by $-0.008$, and a pessimistic $1.4\,\kpckms$ moves it by $+0.050$, smaller than the quoted uncertainty. Correcting for either leaves the exponent inside the range a midplane layer predicts, so a hotter birth does not change which geometry it selects. The larger shift is upward, so the measurement would overstate the exponent rather than understate it.

The cloud sequence is treated as Markovian, which requires successive kicks to be independent. Individual passages last from a few Myr at compact impact parameters to about $30$\,Myr near the adopted outer cutoff, against a vertical period of about $80$\,Myr, so the ordering holds for clouds. It fails for a long-lived spiral wave or bar resonance whose phase stays locked to the star, and \citet{antoja2018} place a coherent perturbation between $300$ and $900$\,Myr ago. Our cut at $\tau>2$\,Gyr removes the stars where that signature is clearest, but every star left in the sample was already present when the perturbation happened, so the cut is not protection against it. Section~\ref{sec:mixture} bounds a coherent component directly instead.

Equation~\eqref{eq:master} also treats encounters as local and independent and takes the perturbing field as externally imposed. It therefore omits the wake raised by a passing cloud \citep{juliantoomre1966} and dressed resonant exchanges between orbits that never come close \citep{fouvry2017}. These form a collective channel absent from the local close-encounter calculation of \citet{spitzer1953}. Our measurement constrains the close-encounter mechanism and is silent about the resonant one, but resonant diffusion is not organised by height above the plane, so it has no reason to imitate the semicircular window. It would add a component with its own action dependence, which the mixture bound of Section~\ref{sec:mixture} limits.

The amplitude carries approximations the exponent does not. The impulse treatment and the truncation of the Kramers--Moyal series after two moments both fail for a close penetrating encounter, which is also why the Coulomb logarithm needs an inner cutoff, and \citet{ida1993} and \citet{sellwood2008} find in addition that diffusion in a flattened disc is anisotropic and not purely local, so the adopted $\ln\Lambda=3$ is a convention within a local model rather than a measured quantity. The relation $E_R=\kappa\JR$ holds in the epicyclic approximation for near-circular orbits and enters the inferred mass linearly through $V$ \citep{binneytremaine2008}. The archived products supply point estimates rather than the original posteriors, so the error model is imposed globally rather than star by star. Radial migration changes a star's guiding radius while adding little random energy \citep{frankel2020}, so it does not heat the star, though it enters through the birth radii that set $R_{\rm eff}$ and means the scatterers a star met were not only those of its present annulus.

The deeper limitation is that a star has two actions and a cloud encounter changes both. The complete description is a $2\times2$ diffusion tensor and we use one element of it as a scalar, which the non-zero cross coefficient of \citet{binneylacey1988} shows that averaging over $\JR$ does not by itself justify. The calculation also places the clouds at rest, whereas real clouds move with non-circular velocities below $10\,\kms$ \citep{reid2019} and a cloud-to-cloud dispersion of about $8\,\kms$ \citep{starkbrand1989}. Of these the off-diagonal coefficient is the larger omission, being the channel through which in-plane energy becomes vertical energy, and \citet{sellwood2014} notes that clouds redirect a fraction of the random energy that way. Whatever reaches the vertical direction by that route is charged here to the clouds, and therefore to a larger $\Meff$, so the quantity measured is an effective scatterer mass. \citet{hamilton2024} approach the same transport problem from the radial side, where the small observed ratio of heating to migration already constrains which mechanisms can be responsible.

\section{Conclusion}
\label{sec:conclusion}

Where the scatterers sit decides whether the heating carries a scale of its own. A kick changes the action in proportion to the speed at which the star arrives, and the time spent at a given height goes as the inverse of that speed, so together they make a semicircular window on the scattering profile. Scatterers that fill the volume give the orbit no length to be measured against, the heating keeps pace as the orbit grows, and the distribution of vertical action within a coeval population stays exponential. A layer at the midplane supplies a length. Once a star has climbed past that layer, it spends proportionally less of its time inside, the heating falls behind the orbit, and the tail is cut off more sharply.

\begin{itemize}
\setlength{\itemsep}{2pt}
\item Scatterers filling the volume give a diffusivity in vertical action $D\propto J_z$ and an exponential distribution. Scatterers in a midplane layer give $D\propto J_z^{1/2}$ and a three-halves stretched exponential. In general a coeval population has $p(J_z)\propto\exp[-(J_z/J_0)^{2-b}]$, with $b$ the logarithmic slope of the diffusivity and $J_0$ the scale of the distribution, so the tail exponent measures the geometry while the amplitude of the heating stays in $J_0$.
\item Measured on $7589$ low-$\alpha$ red clump stars from APOGEE with \textit{Gaia} astrometry, the scale left free at every age, $b=0.51^{+0.06}_{-0.07}$. A molecular layer of the measured thickness predicts $0.44$--$0.54$ once its flaring, the encounter-speed dependence and a realistic vertical potential are all included. Scatterers filling the volume predict $0.85$--$1.00$ under the same treatment.
\item Injecting each geometry into synthetic catalogues built on the same stars, cuts, selection function and error model recovers both unbiased and separated by $5.6\sigma$. Varying the radial range, the age cut and the assumed action error moves the exponent across a half-range of $0.043$, inside the estimator's own scatter.
\item Running the heating clock backwards on those same free scales gives the product of scatterer surface density and effective mass without using cloud data or fitting anything further. Dividing by the observed molecular surface density gives an effective scatterer mass of $2.7\times10^{6}\,\Msun$ from the combined $2$--$4$\,Gyr sample. Cloud catalogues and mass functions give $1.4$--$3.4\times10^{6}\,\Msun$ for the same quantity, a factor of $2.5$ among themselves, and the stellar value falls inside that range. Older bins appear to return up to $4.8\times10^{6}\,\Msun$ when they too are divided by today's gas density. Integrating the literature star-formation history instead brings all four to $1.9$--$2.8\times10^{6}\,\Msun$, inside the cloud range.
\item Bending waves, dark substructure and the $10^{6}\,\Msun$ black-hole haloes once proposed for this heating act wherever the star is, so every one of them gives $b=1$. A mass can be chosen to make any of them match the amplitude, so the amplitude cannot exclude them. The exponent does not depend on mass at all, and it excludes them as the main agent.
\item A sample without ages returns $b\simeq1$ where the truth is one half, $1.03$ for a constant star-formation rate, because a mixture of scales has a heavier tail than any of its members. That bias spans the whole factor of two between the two geometries, so an age-blind sample returns the volume-filling answer whatever the truth. The classical $b=1$ is thereby explained rather than contradicted, and the same holds for any exponent read from velocity distributions without ages.
\item The scales grow with stellar age $\tau$ as $J_0\propto\tau^{1.12}$, where a constant rate would give $\tau^{0.67}$. The disc heated faster in the past than a fixed rate allows, though the scales alone do not separate that from older stars having been born hotter. That does not carry over to the exponent, which a hotter birth shifts by less than its uncertainty.
\end{itemize}

Shape and amplitude are read from the same fit and point to the same scatterers. The disc is heated near the plane, by objects of giant-molecular-cloud mass.

A volume-filling component is held to a small fraction of the diffusive heating once its partner is required to remain a layer. The effective mass bounds what the clouds supply rather than measuring it, since whatever reaches the vertical direction from in-plane motion is charged to them.

The separation of amplitude from shape is not specific to the vertical direction. An amplitude counts perturbers. A shape records where along the orbit they act. Any secular process written as diffusion in action can be asked the same question, provided the sample is resolved in age and the actions give the shape at each age rather than one of its moments. Surveys that measure ages and actions together therefore hold more than a heating rate. They hold the mechanism behind it.

\begin{acknowledgments}
YST acknowledges a Humboldt Research Fellowship from the Alexander von Humboldt Foundation, held at the Max Planck Institute for Astronomy. Solving a Fokker--Planck equation on German funding, at an institute named after the second of those two, seemed the least that could be offered in return.

The project began over cookie breaks at the Institute for Advanced Study in Princeton in 2019, once \citet{tingrix2019} was finished, in conversations with Jing Luan, then a pen-and-paper theorist. The arrangement was that YST would explain machine learning and Jing would explain the impulse sum of Appendix~\ref{app:impulse} and the effective scatterer mass it leads to in Section~\ref{sec:amplitude}. Jing got much the better end of that trade. She is now a machine learning engineer, sensible enough not to do vertical heating for a living, while YST was left holding the effective scatterer mass.

For several years the question stayed there, about what the scatterers weigh. It took a personal emergency, and the unwelcome quantity of free time that came with it, to ask where they sit instead. The first attempt aimed at higher-order velocity moments, and the large language models working alongside it followed that idea a long way down a blind alley. It was the human who noticed that the whole distribution might be available rather than a few of its moments, which is what made the problem tractable. The Fokker--Planck solution the language models produced along the way was correct but rested on more assumptions than it needed. Recasting it as a ladder between moments was also the human's doing, borrowed from the raising and lowering operators of undergraduate quantum mechanics. That material was last studied under the firm impression that it would never be of use for anything. An invitation to lecture on the subject to the Rix group, which YST had issued to himself, forced the derivation into written notes, and those notes became the template for this paper. The calculation and its physical interpretation are both better for the discussions with HWR that followed.

The physics, the derivations, the structure of the argument and every judgement about what this paper claims are the author's. Claude Opus 5.0 and GPT 5.6 helped turn the lecture notes into prose, revised extensively against the author's corrections, and ran the literature searches that turned up the work of the 1960s to the early 1990s on this problem. That literature was unknown to the author while the calculation was being developed, and the results that bear on it sit inside those papers rather than in their abstracts, so what they contain is not evident from a search. Reading them sharpened the argument in several places.
\end{acknowledgments}

\appendix
\renewcommand{\theHequation}{\Alph{section}.\arabic{equation}}

\section{From the Master Equation to Fokker--Planck}
\label{app:fp}

A Fokker--Planck equation is the small-jump limit of an exact statement about where probability goes. Taking that limit is what fixes which moments of the kick survive and which do not. The exact statement comes first. Let $p(\upsilon,t)$ be the density of some quantity $\upsilon$ driven by an external random field. Over a short interval $\Delta t$ a star at $\upsilon-\zeta$ jumps by $\zeta$ and arrives at $\upsilon$. Writing $K_{\Delta t}(\zeta\mid \upsilon_0)$ for the density of a jump $\zeta$ starting from $\upsilon_0$, normalised so that $\int K\,\dd\zeta=1$, and adding up every way of arriving at $\upsilon$,
\begin{equation}
p(\upsilon,t+\Delta t)=\int p(\upsilon-\zeta,t)\,K_{\Delta t}(\zeta\mid \upsilon-\zeta)\,\dd\zeta .
\label{eq:master-eq}
\end{equation}
One physical assumption enters here, that the jump made in this interval does not remember the jump made in the last one, which is what makes the process Markovian.

To turn equation~\eqref{eq:master-eq} into a differential equation, use the fact that the jumps are small. Both $p$ and $K$ are evaluated at $\upsilon-\zeta$, so Taylor-expanding the product about $\upsilon$ and integrating term by term gives
\begin{equation}
p(\upsilon,t+\Delta t)=\sum_{n=0}^{\infty}\frac{(-1)^{n}}{n!}
\frac{\partial^{n}}{\partial \upsilon^{n}}
\Big[p(\upsilon,t)\,\big\langle\zeta^{n}\big\rangle_\upsilon\Big] ,
\label{eq:km}
\end{equation}
an exact rewriting as an infinite series in the jump moments, known as the Kramers--Moyal expansion. Its $n=0$ term is $p(\upsilon,t)$ itself. Subtracting it, dividing by $\Delta t$ and keeping the two moments that survive the limit defines
\begin{equation}
\mathcal{A}(\upsilon)=\lim_{\Delta t\to0}\frac{\langle\zeta\rangle_\upsilon}{\Delta t} ,
\qquad
D(\upsilon)=\lim_{\Delta t\to0}\frac{\langle\zeta^{2}\rangle_\upsilon}{\Delta t} ,
\end{equation}
and leaves the Fokker--Planck equation
\begin{equation}
\frac{\partial p}{\partial t}
=-\frac{\partial}{\partial \upsilon}\big[\mathcal{A}\,p\big]
+\frac12\frac{\partial^{2}}{\partial \upsilon^{2}}\big[D\,p\big] .
\end{equation}
Higher moments are dropped because for small kicks $\langle\zeta^{n}\rangle$ falls off faster than $\Delta t$, which is legitimate exactly when individual jumps are small compared with $\upsilon$ itself.

If the kicks have no preferred direction and the same strength everywhere, then $\mathcal{A}=0$, $D=D_0$, and this is the heat equation $\partial_tp=\tfrac12D_0\partial_\upsilon^2p$, whose solution from a point is the spreading Gaussian. Section~\ref{sec:theory} solves that case through its moments and then repeats the move on the half-line, where the diffusivity is no longer constant and the moments belong to a less familiar distribution.

\section{The Similarity Reduction}
\label{app:similarity}

Section~\ref{sec:theory} obtains equation~\eqref{eq:solution} from the moments. The same result follows by reducing the partial differential equation itself, which confirms it and is the route \citet{binneylacey1988} took for the case they treated.

Equation~\eqref{eq:fp} carries no scale of its own. Here $D_0J_z^{\,b}$ is a pure power, and a population born cold starts from a point. There is therefore no action against which to measure $J_z$ except one built from the problem, so the distribution can only keep a fixed shape and stretch. Writing that as
\begin{equation}
p(J_z,t)=\frac{1}{J_0(t)}\,F(x) ,\qquad x\equiv\frac{J_z}{J_0(t)} ,
\label{eq:ansatz}
\end{equation}
with the prefactor keeping the normalisation independent of $t$, the left-hand side of equation~\eqref{eq:fp} becomes $-J_0^{-2}(\dd J_0/\dd t)(F+xF')$ and the right-hand side $\tfrac12D_0J_0^{\,b-3}(x^{b}F')'$, a prime denoting $\dd/\dd x$. Neither side may depend on $t$ except through an overall factor, so the two agree for all $t$ only if $\dd J_0/\dd t\propto J_0^{\,b-1}$. Fixing the constant, which amounts to choosing the units of $x$, gives $\dd J_0/\dd t=\tfrac12(2-b)D_0J_0^{\,b-1}$, and integrating from $J_0(0)=0$ returns equation~\eqref{eq:clock}. The scale therefore follows from the matching rather than having to be supplied.

That leaves an ordinary differential equation in $x$ alone,
\begin{equation}
\big(x^{b}F'\big)'+(2-b)\,\big(xF\big)'=0 .
\label{eq:shape}
\end{equation}
Both terms are exact derivatives, so this integrates once without effort. The constant of integration is the flux of stars through $J_z=0$, and it vanishes because an action cannot be negative and no stars are created or destroyed at the origin. That leaves $x^{b}F'=-(2-b)xF$, or $\dd\ln F/\dd x=-(2-b)x^{\,1-b}$, whose integral is $\ln F=-x^{2-b}$ up to a constant. Normalising with $\int_0^\infty\exp(-x^{2-b})\,\dd x=\Gamma(\tfrac{1}{2-b})/(2-b)$ recovers equation~\eqref{eq:solution}.

\citet{binneylacey1988} obtained three cases. First, they found $b=1$ by requiring that a Gaussian velocity distribution stay Gaussian, which gave $D_{zz}\propto E_z$ and $\sigma\propto\sqrt{t}$. Second, they found $b=0$, a distribution Gaussian in the energies with $\sigma\propto t^{1/4}$, for a thin cloud layer in three dimensions when the star's own motion sets the encounter speed. Third, they found $b=1/2$ as the late-time solution for a razor-thin disc,
\begin{equation}
f\propto t^{-2/3}\exp\!\left[-\frac{\big(v_R^{2}+\gamma^{2}v_\phi^{2}\big)^{3/2}}
{C\,t}\right] ,
\end{equation}
in which $v_R$ and $v_\phi$ are the in-plane velocity components, $\gamma$ is the axis ratio of the epicycle, so that $v_R^{2}+\gamma^{2}v_\phi^{2}=2E_R$ is twice the epicyclic energy, and $C$ collects the constants of their cloud population. Since $E_R\simeq\kappa\JR$, that is equation~\eqref{eq:solution} written in the radial action, with their similarity variable $\xi^{3/2}/\tau$ playing the part of our $J_z^{2-b}/t$.

\citet{binneylacey1988} took this route rather than the moments. Their reduction rests on the same condition as the one above, the branch of the solution that carries no flux through the origin, and differs in two ways. Their similarity variable is fixed in advance as $\xi^{3/2}/\tau$, its exponent following from a diffusivity already known to be $\sqrt{\xi}$. Leaving $J_0(t)$ undetermined instead makes equation~\eqref{eq:clock} a consequence of the matching. Their reduced equation is also of second order, solved by a power series from which the exponential emerges only for the one star-formation history that keeps the number of stars fixed. The normalisation-preserving prefactor of equation~\eqref{eq:ansatz} makes both terms of equation~\eqref{eq:shape} exact derivatives instead, so it integrates by inspection and for every $b$ at once.

The moment route of Section~\ref{sec:theory} needs one thing the similarity reduction does not, a determinate moment problem, which for a tail $\exp(-x^{2-b})$ on the half-line holds for $2-b\ge1/2$. The measured $2-b\simeq1.5$ is well inside that.

\section{Finite Layer Thickness and the Encounter-Speed Dependence}
\label{app:thickness}

Two idealisations enter the prediction $b=1/2$, and they act in opposite directions. The first is that the layer has zero thickness. For a Gaussian profile of scale height $h$ normalised so that $\int q\,\dd z=Q$, equation~\eqref{eq:master} gives a diffusivity whose local slope interpolates between the two ideal cases, as Figure~\ref{fig:profiles} shows. The slope is meaningful only locally, because the window radius $Z$ grows with action while $h$ does not. The entries of Table~\ref{tab:geometry} average it over the action distribution the selection function is corrected back to, which is the population the theory describes. The local slope runs from $0.50$ at large action to above $0.7$ near the peak, so the table reports that average rather than any one value of it.

The second idealisation is that the encounter speed does not depend on the vertical action. It does. Writing the scattering profile as its column times a unit-normalised shape, $q(z)=Q\,\hat g(z)$, equation~\eqref{eq:master} factorises exactly,
\begin{equation}
D(J_z)=Q\times\frac{1}{\pi}\int_{-Z}^{Z}\hat g(z)\sqrt{Z^{2}-z^{2}}\;\dd z ,
\end{equation}
so the geometry sits entirely in the second factor and the encounter speed entirely in the first, through the $Q\propto1/V$ of equation~\eqref{eq:Q}. Taking logarithms turns the product into a sum,
\begin{equation}
b=\frac{\dd\ln(\text{geometry})}{\dd\ln J_z}-\frac{\dd\ln V}{\dd\ln J_z} ,
\end{equation}
which is why the two corrections add rather than interfere. With $V^{2}\simeq2\kappa\JR+2\nu J_z=2E_R+2E_z$, differentiating at fixed $\JR$ gives
\begin{equation}
\frac{\dd\ln V}{\dd\ln J_z}=\frac{\nu J_z}{V^{2}}
=\frac{1}{2}\,\frac{1}{1+E_R/E_z} .
\end{equation}
The term vanishes when the in-plane motion dominates and reaches $1/2$ when the motion is mostly vertical, in which case it cancels the geometric slope of a sheet entirely.

Neither limit describes the sample. Both actions come from \citet{tingrix2019}, and turning them into energies takes the two frequencies, for which we adopt $\nu=75\,\kms\,{\rm kpc}^{-1}$ and $\kappa=\sqrt{2}\,v_c/R_\odot=41\,\kms\,{\rm kpc}^{-1}$ for every star, since the archived products do not resolve either star by star. The $E_R/E_z$ that follows has percentiles $(0.60,\,2.86,\,13.77)$, so the in-plane motion is the larger contributor for most stars but not overwhelmingly so, and the correction is neither zero nor one half. Averaged over the population the term is $0.163$, and scaling $\kappa/\nu$ by $\pm30$ per cent moves it by only $\mp0.025$.

\begin{table}[t]
\centering
\caption{The prediction for a real cloud layer, against the volume-filling case treated the same way. Thickness is the mass-weighted rms height in the cloud catalogue, and the layer flares, so the inner and outer annuli are shown alongside the sample. The geometric slope is the local $\dd\ln D/\dd\ln J_z$ of equation~\eqref{eq:master} averaged over the action distribution the selection is corrected back to, and the encounter-speed term is $\langle\tfrac12(1+E_R/E_z)^{-1}\rangle$ over the same stars. The harmonic row is the sum of the two above it, and the predicted range adds the anharmonic shift to that. Volume-filling scatterers have no layer to thicken, so their geometric slope is $1$ by construction.}
\label{tab:geometry}
\small
\setlength{\tabcolsep}{4.0pt}
\begin{center}
\hspace*{-53pt}\begin{tabular}{lcccc|c}
\toprule
& \multicolumn{4}{c|}{Midplane layer} & Volume\\
& Sheet & $5$--$7$\,kpc & $5$--$10$\,kpc & $9$--$10$\,kpc & filling\\
\midrule
Catalogue-derived $h$ (pc) & $0$ & $81$ & $102$ & $145$ & --\\
Geometry, $b_{\rm geom}$ & $0.500$ & $0.569$ & $0.596$ & $0.654$ & $1.000$\\
Encounter speed & $-0.163$ & $-0.163$ & $-0.163$ & $-0.163$ & $-0.163$\\
\midrule
Harmonic $b$ & $0.337$ & $0.406$ & $0.433$ & $0.491$ & $0.837$\\
Anharmonic shift & \multicolumn{4}{c|}{$+0.01$ to $+0.11$} & $+0.01$ to $+0.16$\\
\midrule
Predicted $b$ & $0.35$--$0.45$ & $0.42$--$0.52$ & $0.44$--$0.54$ & $0.50$--$0.60$ & $0.85$--$1.00$\\
\bottomrule
\end{tabular}
\end{center}
\end{table}

The two therefore move the exponent by comparable amounts and in opposite senses. The encounter speed is the larger, so the net is a shift downward from the ideal sheet, and neither is large enough to carry the prediction to $1$. One idealisation is left, the harmonic potential, and relaxing it pushes back the other way.

\section{Anharmonic Vertical Potentials}
\label{app:anharmonic}

The real vertical force steepens near the plane and flattens away from it. Two things could be lost when the quadratic potential is dropped, the form of the equation and the value of the exponent. The first survives exactly and only the second moves.

The harmonic approximation enters only in the map from $q(z)$ to $D(J_z)$, where it supplies the closed-form window $\sqrt{Z^{2}-z^{2}}$. To see that the rest does not depend on it, run the argument again without writing down an orbit. Work first in velocity at a fixed height, where the perturbers deliver zero mean kick and a mean square kick $q(z)$ per unit time, so that
\begin{equation}
\frac{\partial f}{\partial t}
=\frac12\frac{\partial}{\partial v_z}\!
\left[q(z)\frac{\partial f}{\partial v_z}\right] .
\end{equation}
No drift has to be added, because $q$ depends on height alone. A drift would require the kick strength to change with the star's own velocity, and a passing cloud does not know how fast the star is moving.

Now change to action and angle. The transformation is canonical, so $\dd z\,\dd v_z=\dd J_z\,\dd\theta$ and a phase-mixed population is $f=p(J_z,t)/2\pi$. Every one-dimensional vertical potential has an orbital frequency $\Omega(J_z)$ obeying $\dd J_z/\dd E_z=1/\Omega$, and $E_z$ carries the velocity as $\tfrac12v_z^{2}$, so $\partial J_z/\partial v_z=v_z/\Omega(J_z)$. A flux transforms with that same factor. Averaging around the orbit, the part that merely carries stars around in phase cancels, and what is left is
\begin{equation}
\frac{\partial p}{\partial t}
=\frac12\frac{\partial}{\partial J_z}\left[D\frac{\partial p}{\partial J_z}\right] ,
\qquad
D(J_z)=\frac{\big\langle q\,v_z^{2}\big\rangle_\theta}{\Omega^{2}(J_z)} ,
\end{equation}
which is the harmonic result with $\nu$ replaced by $\Omega(J_z)$, and nothing else changed. The conservative form, and with it the relation between drift and diffusion, follows from phase mixing and from kicks that are symmetric and depend on position alone, not from the quadratic potential.

Writing that diffusivity out needs one more identity. The dwell-time average with $\oint v_z\,\dd z=2\pi J_z$ gives $\langle v_z^{2}\rangle_t=\Omega J_z$, and with it the orbit average becomes
\begin{equation}
D(J_z)=\frac{1}{\pi\,\Omega(J_z)}\int q(z)\,\big|v_z(z;J_z)\big|\;\dd z ,
\end{equation}
with $D=q_0J_z/\Omega$ for scatterers filling the volume and $D=Qv_0/\pi\Omega$ for a sheet, where $v_0$ is the midplane speed. Because $\Omega$ falls with $J_z$ in a flattening potential, both slopes are pushed up.

To measure by how much, we take a self-gravitating disc softened over a scale height $h_\star$ inside a harmonic halo. The potential is $\Phi(z)=2\pi G\Sigma_\star(\sqrt{z^2+h_\star^2}-h_\star)+\tfrac12\nu_{\rm h}^2z^2$, quadratic near the plane and tending to the linear potential of a sheet far from it. We use $\Sigma_\star=45$, $50$ and $55\,\Msun\,{\rm pc}^{-2}$ with $h_\star=0.25$, $0.30$ and $0.45$\,kpc, in each case fixing $\nu_{\rm h}$ so that the total midplane frequency is the $75\,\kms\,{\rm kpc}^{-1}$ used throughout, which leaves $\nu_{\rm h}=28$, $33$ and $48\,\kms\,{\rm kpc}^{-1}$. Solving each orbit for $\Omega(J_z)$ and evaluating both diffusivities at the observed actions gives $b=1.01$--$1.16$ for volume filling and $0.51$--$0.61$ for a sheet. Both move up, and they move together. The distance between them stays within $0.50$--$0.56$ against the harmonic $0.5$, so the separation the measurement relies on survives. Table~\ref{tab:geometry} collects the three corrections, and for the measured thickness they predict $0.44$--$0.54$.

\section{The Selection-Corrected Normalisation}
\label{app:likelihood}

Multiplying the model by the selection function takes away its normalisation, so each component of equation~\eqref{eq:likelihood} has to be renormalised over the range the survey can see. Equation~\eqref{eq:norm} is that renormalisation, and it has to be redone for every star at every trial parameter set, which a numerical quadrature inside the likelihood would make prohibitive. The two-branch form of equation~\eqref{eq:selection} avoids that, because it leaves the integral in closed form. The same selection-corrected normalisation was developed in \citet{tingrix2019}.

Write $s\equiv2-b$ for the tail exponent and normalise the solution of equation~\eqref{eq:solution},
\begin{equation}
p(J_z\mid J_0)=\frac{s}{J_0\,\Gamma(1/s)}\exp\!\left[-\left(\frac{J_z}{J_0}\right)^{s}\right] ,
\end{equation}
which follows from $\int_0^\infty\exp(-x^{s})\,\dd x=\Gamma(1/s)/s$. Substituting $u=(J_z/J_0)^{s}$, so that $J_z=J_0u^{1/s}$ and $\dd J_z=(J_0/s)\,u^{1/s-1}\dd u$, turns every integral of $p$ against a power of $J_z$ into an incomplete gamma function.

Below the transition the selection is unity, and the substitution gives
\begin{equation}
\int_0^{J_{{\rm t},i}}\!p\,\dd J_z
=\frac{1}{\Gamma(1/s)}\int_0^{u_{{\rm t},i}}\!e^{-u}u^{1/s-1}\,\dd u
=\frac{\gamma\!\left(1/s,\,u_{{\rm t},i}\right)}{\Gamma(1/s)} ,
\end{equation}
with $u_{{\rm t},i}=(J_{{\rm t},i}/J_0)^{s}$. Above it the integrand carries the extra factor $c_iJ_z^{-\xi_i}=c_iJ_0^{-\xi_i}u^{-\xi_i/s}$, which shifts the exponent of $u$ by $-\xi_i/s$ and leaves the same function,
\begin{equation}
\int_{J_{{\rm t},i}}^{J_{\max}}\!\!c_iJ_z^{-\xi_i}\,p\,\dd J_z
=\frac{c_iJ_0^{-\xi_i}}{\Gamma(1/s)}
\left[\Gamma\!\left(\tfrac{1-\xi_i}{s},u_{{\rm t},i}\right)
-\Gamma\!\left(\tfrac{1-\xi_i}{s},u_{\max}\right)\right] ,
\end{equation}
where $\gamma$ and $\Gamma$ are the lower and upper incomplete gamma functions and $u_{\max}=(J_{\max}/J_0)^{s}$ with $J_{\max}=1000\,\kpckms$. Adding the two branches,
\begin{equation}
Z_{ik}\;\propto\;
\gamma\!\left(\tfrac{1}{s},u_{{\rm t},i}\right)
+c_i\,J_0^{-\xi_i}\left[\Gamma\!\left(\tfrac{1-\xi_i}{s},u_{{\rm t},i}\right)
-\Gamma\!\left(\tfrac{1-\xi_i}{s},u_{\max}\right)\right] .
\end{equation}

Both terms depend on $b$ through the first argument of the incomplete gamma function, so the selection correction moves every time the exponent does. The three parameters vary from star to star with radius, so $J_{{\rm t},i}$, $\xi_i$ and $c_i$ change while the closed form above does not.

\section{Validation by Injection and Recovery}
\label{app:validation}

The exponent is read from the tail of a distribution whose scale is itself free at four ages, and the selection function has already thinned that tail. Freedom taken at the low-action end could in principle be paid for at the high end, where the exponent lives, and nothing in the fit itself would reveal it. Nor can the fit say whether the width of its own likelihood is the width the estimator has from one sample to the next. Both questions are answered by giving the estimator data whose answer is known.

Each synthetic catalogue keeps the real stars and replaces only their actions. A star of true age $\tau$ and radius $R_{\rm eff}$ is given a scale $J_0$ from the history the real fit returned, the four knot values and the radial slope of Section~\ref{sec:exponentresult}, and its action is drawn from equation~\eqref{eq:solution} at the chosen $b$, which for a cold start means $(J_z/J_0)^{2-b}$ is a unit gamma deviate of shape $1/(2-b)$. The draw is then accepted with the probability equation~\eqref{eq:selection} assigns it at that star's own radius, and only after that is the $20$ per cent lognormal action error applied. Selection therefore acts on the true action and the error afterwards, in the order the sky imposes them.

The age cut needs the same care. It is applied to the observed age, so the stars in the real fit have true ages spread either side of the window. Drawing from a parent pool of $0.5$--$13$\,Gyr, scattering by the $25$ per cent age error and only then cutting at $2$--$8$\,Gyr reproduces that spread. Six trials are run at each geometry.

\begin{figure}[t]
\centering
\includegraphics[width=0.62\columnwidth]{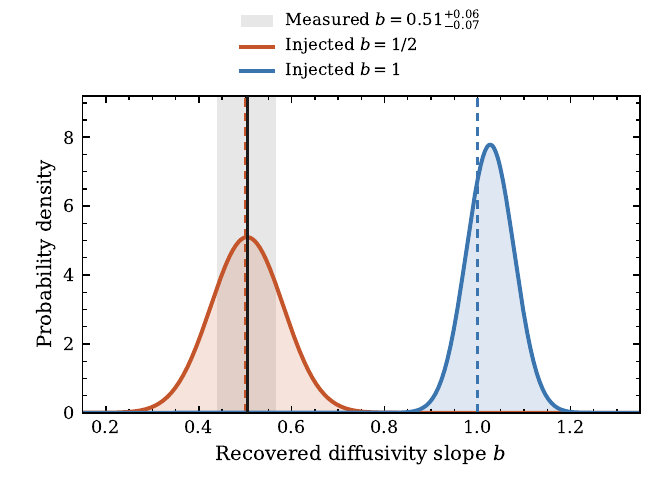}
\caption{Exponents recovered from synthetic catalogues whose true $b$ is known. Each catalogue keeps the real stars, cuts, selection function and error model and replaces only the actions, drawing them at the injected $b$ from the scale history the fit of Section~\ref{sec:exponentresult} returned. Six trials are run at each geometry, and each curve is the Gaussian of the mean and scatter they return, with the dashed line marking the value injected. A midplane layer returns $0.504\pm0.078$ and volume filling $1.027\pm0.051$, unbiased in both cases and $5.6\sigma$ apart. The black line and grey band are the measurement, $b=0.51^{+0.06}_{-0.07}$, which the injected midplane value sits inside.}
\label{fig:recovery}
\end{figure}

Injecting $b=0.500$ returns $0.504\pm0.078$ and injecting $b=1.000$ returns $1.027\pm0.051$, unbiased in both cases and separated by $5.6\sigma$, as Figure~\ref{fig:recovery} shows. The scatter is a little wider than the posterior interval of Section~\ref{sec:exponentresult}, and it is smaller at $b=1$, where a steeper diffusivity leaves a sharper tail to measure. The separation is wide enough to tell the two geometries apart on data of this quality.

The same machinery tests an assumption the real fit makes. Equation~\eqref{eq:solution} is the solution for a population born cold, and Section~\ref{sec:measurement} argued that a birth distribution of the same shape costs nothing, since it only adds a constant to $J_0^{\,2-b}$ and the free scales absorb it. A birth distribution concentrated at a single action is not of that shape and is therefore the least favourable case, so that is what we inject.

We draw it from the Green function of equation~\eqref{eq:fp} for a source at $J_{\rm b}$, holding the birth term and the heating so that their sum is the $J_0^{\,2-b}$ the fit already assigns. The injected population then carries the scale the data carry, and differs from a cold start only in shape, which is what the exponent is read from. Injecting $b=0.500$ then returns $0.496\pm0.037$ for $J_{\rm b}=0.65\,\kpckms$, the value $\sigma_{\rm gas}^{2}/\nu$ gives for a $7\,\kms$ gas layer, and $0.554\pm0.020$ for a pessimistic $1.4\,\kpckms$, against $0.504\pm0.078$ for the cold start. The realistic birth action moves the exponent down by $0.008$ and the pessimistic one up by $0.050$, and six trials resolve neither against the scatter of the cold start itself. Assuming a cold start therefore costs less than the uncertainty already quoted, in either direction.

\section{The Impulse Sum and the Coulomb Logarithm}
\label{app:impulse}

Everything the perturbers contribute reaches the diffusivity through the single column $Q$, so the contents of that column are what the heating amplitude measures. Put closest approach at $t=0$, so that a cloud of mass $M$ passing at impact parameter $\varpi$ with relative speed $V$ is at separation $r=\sqrt{\varpi^{2}+V^{2}t^{2}}$ and pulls on the star with a component $GM\varpi/r^{3}$ perpendicular to its motion. The parallel component is antisymmetric in $t$ and integrates to zero. Substituting $Vt=\varpi\tan\theta$, so that $\dd t=\varpi\sec^{2}\theta\,\dd\theta/V$ and $(\varpi^{2}+V^{2}t^{2})^{3/2}=\varpi^{3}\sec^{3}\theta$, the $\sec^{3}\theta$ cancels all but a cosine,
\begin{equation}
\int_{-\infty}^{\infty}\frac{\dd t}{(\varpi^{2}+V^{2}t^{2})^{3/2}}
=\frac{1}{V\varpi^{2}}\int_{-\pi/2}^{\pi/2}\cos\theta\,\dd\theta
=\frac{2}{V\varpi^{2}} ,
\end{equation}
and therefore
\begin{equation}
\Delta v_\perp=\frac{2GM}{\varpi V} .
\end{equation}
The kick is linear in the mass and inverse in the encounter speed, a faster passage leaving less time for the force to act. Both survive to the end. Linearity in $M$ is what puts a second moment of the mass function into the answer, and the inverse speed is what puts $V$ in the denominator of $Q$.

In a time $\dd t$ the star sweeps a cylindrical shell of radius $\varpi$ and thickness $\dd\varpi$, so encounters within it occur at a rate $\dd\mathcal{R}=2\pi\varpi\,\dd\varpi\,V\,n(z,M)\,\dd M$ for a number density $n(z,M)$. Successive encounters are uncorrelated, so it is their squares that add. The plane perpendicular to the star's motion has two independent directions, and since the relative velocity is mostly in-plane one of them is roughly vertical, so half the perpendicular variance is vertical and
\begin{equation}
\dd q=\tfrac12(\Delta v_\perp)^{2}\,\dd\mathcal{R}
=\frac{4\pi G^{2}}{V}\,M^{2}\,n(z,M)\,\frac{\dd\varpi}{\varpi}\,\dd M .
\end{equation}
Splitting the variance evenly is a scalar stand-in for an average over the velocity ellipsoid, which is the approximation Section~\ref{sec:discussion} returns to.

The impact-parameter dependence has collapsed to $\dd\varpi/\varpi$, so every decade of encounter distance contributes equally, and integrating it gives the Coulomb logarithm $\ln\Lambda=\ln(\varpi_{\max}/\varpi_{\min})$. We take the inner cutoff to be a cloud radius, below which the impulse approximation fails, and adopt $V/(2A)$ as the outer cutoff, with $A$ the Oort constant, where the encounter duration matches the shear time. This shear-scale convention differs from the epicycle-scale outer limit used by \citet{lacey1984}. In the catalogue of Section~\ref{sec:amplitude} the inner cutoff is about $20$\,pc for a typical cloud and about $100$\,pc for the massive complexes that carry most of the second moment, and the outer is one to two kiloparsecs. Averaging $\ln(\varpi_{\max}/\varpi_{\min})$ over those clouds with the $M^{2}$ weight the diffusivity gives them returns $2.8$, or $2.5$ with Lacey's epicycle cutoff, while weighting the clouds equally returns $4.7$. We adopt $\ln\Lambda=3$.

Integrating over height, and writing $N(M)=\int n(z,M)\,\dd z$ for the number per unit area, the column is
\begin{equation}
Q=\frac{4\pi G^{2}\ln\Lambda}{V}\int M^{2}N(M)\,\dd M .
\end{equation}
Splitting that integral into a surface density $\Ssc=\int MN(M)\,\dd M$ and the ratio $\Meff=\int M^{2}N(M)\,\dd M\big/\int MN(M)\,\dd M$ gives equation~\eqref{eq:Q},
\begin{equation}
Q=\frac{4\pi G^{2}\,\Ssc\,\Meff\,\ln\Lambda}{V} .
\end{equation}
The second moment appears because the impulse is linear in $M$ while $q$ is quadratic in it, so which moment the heating responds to is fixed by the derivation. The measurement returns its value through the product $\Ssc\Meff$.

\section{The Exponent Recovered from an Age-Blind Sample}
\label{app:blind}

\begin{figure}[t]
\centering
\includegraphics[width=0.5\columnwidth]{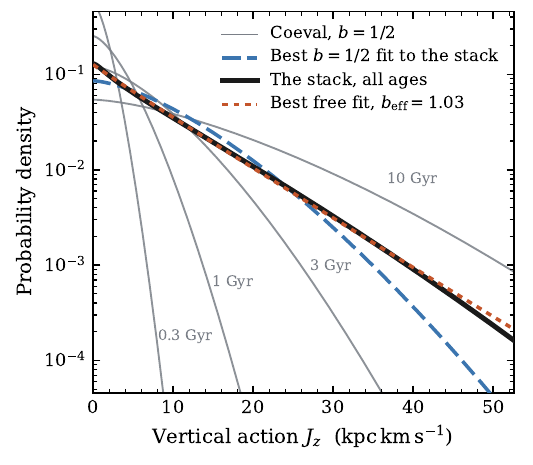}
\caption{The shape an age-blind sample returns, against the populations it is built from. The grey curves are coeval populations with $b=1/2$ at four ages, differing only in the scale the clock assigns them, and each bends downwards on a logarithmic axis as a three-halves stretched exponential must. The clock is normalised through the measured $J_0$ at $5$\,Gyr so that the axis carries physical units, a choice the exponent does not depend on. Their stack over $0.1$--$10$\,Gyr at a constant formation rate, across which the scale grows twenty-fold, is close to straight, which is what $b=1$ looks like. Both fitted curves maximise equation~\eqref{eq:crossentropy} against the stack, one with the exponent free and one with it held at the true $3/2$. The free fit returns $b_{\rm eff}=1.03$ and holds the stack to within a per cent at its $99$th percentile in $J_z$, where holding the true $b=1/2$ instead falls short by a factor of two.}
\label{fig:ageblind}
\end{figure}

The measurement of Section~\ref{sec:measurement} uses an age for every star. Without ages the same stars give a different answer. Stars of different ages carry different scales, and far out in action only the oldest and widest still have stars in them, so the superposition falls off more slowly than any population it is built from, as Figure~\ref{fig:ageblind} shows. A slower fall-off is read as a larger exponent. A population that truly has $b=1/2$ is then fitted by a value near unity, which is what volume-filling scatterers would give, so an age-blind sample cannot draw the distinction at all.

Age enters the theory only through the scale, so the observed distribution is a superposition of similarity solutions and its moments are the coeval moments averaged over age. Integrating equation~\eqref{eq:solution} against a power of the action, each coeval population contributes $\langle J_z^{\,n}\rangle=J_0(\tau)^{n}\, \Gamma\!\left(\tfrac{n+1}{s}\right)/\Gamma\!\left(\tfrac1s\right)$ with $s\equiv2-b$, of which equation~\eqref{eq:moments} is the case $n=ms$. The clock of equation~\eqref{eq:clock} supplies $J_0\propto\tau^{1/s}$. Weighting by a star-formation rate $\propto\tau^{\,\eta}$ between $\tau_{\min}$ and $\tau_{\max}$,
\begin{equation}
\big\langle J_z^{\,n}\big\rangle_{\rm mix}
=\frac{\Gamma\!\left(\frac{n+1}{s}\right)}{\Gamma\!\left(\frac1s\right)}
\cdot\frac{\displaystyle\int_{\tau_{\min}}^{\tau_{\max}}
\tau^{\,n/s+\eta}\,\dd\tau}
{\displaystyle\int_{\tau_{\min}}^{\tau_{\max}}\tau^{\,\eta}\,\dd\tau} .
\label{eq:mixmoment}
\end{equation}

Fitting that mixture with a single member of the family assigns it one pair $(J_0,s')$. In the limit of a large sample, maximising the likelihood becomes maximising the mean log-likelihood per star under the true distribution. Writing that mean out,
\begin{equation}
\big\langle\ln p\big\rangle=\ln s'-\ln\Gamma\!\left(\tfrac{1}{s'}\right)-\ln J_0
-\frac{\big\langle J_z^{\,s'}\big\rangle_{\rm mix}}{J_0^{\,s'}} ,
\label{eq:crossentropy}
\end{equation}
the mixture appears in one place only. Setting $\partial\langle\ln p\rangle/\partial J_0=0$ gives $J_0^{\,s'}=s'\langle J_z^{\,s'}\rangle_{\rm mix}$, the scale at which the fitted shape reproduces that moment, and putting it back leaves a function of $s'$ alone,
\begin{equation}
\ln s'-\ln\Gamma\!\left(\tfrac{1}{s'}\right)
-\frac{1}{s'}\ln\!\left(s'\big\langle J_z^{\,s'}\big\rangle_{\rm mix}\right)
-\frac{1}{s'} ,
\end{equation}
which equation~\eqref{eq:mixmoment} supplies in closed form.

Only the exponent is determined by that expression. An overall factor in $J_0(\tau)$ raises $\langle J_z^{\,s'}\rangle_{\rm mix}$ by that factor to the power $s'$, which the $1/s'$ in front of the logarithm turns into an additive constant, so it cannot move the maximum. Equation~\eqref{eq:mixmoment} is therefore written with that factor set to unity. Maximising over $s'$ gives the exponent that an age-blind fit reports, $b_{\rm eff}=2-s'$.

The comparison is with exponents measured before large age catalogues existed, so the mixture spans the disc, $0.1$--$10\,$Gyr, and no selection function is applied. For a constant rate, $b=1/2$ returns $b_{\rm eff}=1.03$, the case Figure~\ref{fig:ageblind} draws, and $b=1$ returns $1.41$. With $\eta=1$, a rate rising towards early times, the same two return $0.75$ and $1.20$. The displacement is upwards in every case, which is what the tail argument requires. How far it moves depends on the star-formation history, which is why an age-blind exponent cannot be corrected after the fact.

\bibliographystyle{aasjournal}
\bibliography{references}

\end{document}